\documentstyle[12pt,prb,aps]{revtex}
\tiny\normalsize 
\input epsf
\begin{document}
\draft

\title{The profile of a decaying crystalline cone}
\author{Navot Israeli\cite{NavotEmail} and Daniel
Kandel\cite{DanielEmail}}
\address{Department of Physics of Complex Systems,\\
Weizmann Institute of Science, Rehovot 76100, Israel}
\maketitle

\begin{abstract}
The decay of a crystalline cone below the roughening
transition is studied. We consider local mass transport through
surface diffusion, focusing on the two cases of diffusion limited
and attachment-detachment limited step kinetics. In both cases, we
describe the decay kinetics in terms of step flow models.
Numerical simulations of the models indicate that
in the attachment-detachment limited case the system undergoes 
a step bunching instability if the repulsive interactions between steps
are weak. Such an instability does not occur in the
diffusion limited case. In stable cases 
the height profile, $h(r,t)$, is flat at radii $r<R(t)\sim
t^{1/4}$. Outside this flat region the height profile obeys the
scaling scenario $\partial h/\partial r = {\cal F}(r t^{-1/4})$. A
scaling ansatz for the
time-dependent profile of the cone yields analytical values for
the scaling exponents and a differential equation for the scaling
function. In the long time limit this equation provides an exact
description of the discrete step dynamics. It admits 
a family of solutions and the mechanism responsible for the
selection of a unique scaling function is discussed in detail. Finally
we generalize the
model and consider permeable steps by allowing direct adatom hops
between neighboring terraces. We argue that step permeability does not
change the scaling behavior of the system, and its only effect is a
renormalization of some of the parameters.

\end{abstract}

\pacs{68.35.Bs, 68.55.-a,  68.55.Jk}

\section{Introduction}

The properties of crystalline nanostructures are of considerable
interest, because of the technological importance of nanostructures in
fabrication of electronic devices. Kinetic properties of nanostructures
attracted particular attention, since in many cases nanostructures are
thermodynamically unstable and tend to decay with time. Such decay
processes have been studied both theoretically and experimentally.

Considerable effort was devoted to the study of periodic
structures. The decay of one and two dimensional gratings was 
studied extensively. The emerging experimental
picture\cite{Tanaka,Blakely,Umbach,Yamashita}  
is that below
the roughening temperature these
structures decay in a shape preserving manner. Macroscopic facets are
observed at the maxima and minima of the gratings. Although these
systems are out of
equilibrium, the appearance of facets is
a manifestation of the cusp singularity of the surface free energy at
the high symmetry crystalline orientation. 

There are basically two theoretical approaches to the problem of surface
evolution in general and nanostructure decay in particular. On the one
hand, there are phenomenological models which treat the   
crystal surface as a continuous
medium\cite{Mullins,BonzelMullins,BonzelPreuss,HagerSpohn,%
LanconVillain,OzdemirZangwill}.
The evolution of the surface is then driven by the tendency of the
system to lower its free energy (given in terms of continuous spatial
variables). The advantage of such models is that  
they are relatively simple and can sometimes lead to analytical
predictions of surface evolution. However, these  
models ignore the discrete nature of surface steps, which may become
important below the roughening transition.
In addition, most of them rely on assumptions of small surface
slope and/or surface curvature (with the  
exception of Ref.\ \onlinecite{HagerSpohn}), and are unable to properly
treat the behavior of the macroscopic facets observed experimentally.

On the other hand, there are models which treat surface evolution
on a smaller scale.  
Among these are microscopic
models\cite{RamanaCooper,DubsonJeffers,JiangEbner,ErlebacherAziz},
where the basic 
degrees of freedom are individual atoms, and step flow
models\cite{OzdemirZangwill,RettoriVillain,TanakaBartelt,%
DuportChameMullinsVillain,AdamChameLanconVillain}.
These models are usually solved numerically, and provide results 
which can be directly related to the microscopic dynamics. However, in
most cases, it is difficult to understand the behavior of the system on
larger length scales on the basis of these results.

Research efforts were also directed towards isolated surface structures.
The decay of isolated step bunches, islands or hills was studied both
experimentally\cite{FuJohnsonWeeksWilliams,Morgenstern} and  
theoretically\cite{Uwaha,Villain,SelkeDuxbury}. Here too there is an
apparent gap between the microscopic 
and macroscopic theoretical approaches. 

In this work we attempt to bridge over this
gap in the case of a simple surface structure, i.e. an infinite
crystalline cone. We give a
complete account of the surface dynamics based on a step flow
model, and then derive a continuum model which gives a very accurate
description of the evolution of the cone and becomes exact in the long
time limit. We do not make any assumptions of small slope or small
surface curvature. 

The crystalline cone
consists of an infinite number of circular concentric steps. A
similar system was studied by Rettori and Villain\cite{RettoriVillain}
who considered the decay of bidirectional surface modulations. Their
results are relevant in the case of small amplitude modulations when
the profile peaks and valleys affect each other. Our work addresses the
opposite situation when a single peak  
can be considered as an isolated structure and in this sense  
is complimentary to theirs.

Below the roughening transition atomic steps have a finite free energy.
Their existence on the surface strongly affects its morphological
evolution. In many cases, one can ignore the formation of islands and
voids on the surface and consider only adatom diffusion and attachment
and detachment processes to and from step edges. The decay of a
nanostructure is then dominated by the motion of steps. In order to
describe the decay process mathematically, one has to solve the
diffusion equation for adatoms on the terraces between the steps, with
boundary conditions at the step edges, in the spirit of the
Burton-Cabrera-Frank model\cite{BCF}. If the geometry of the
nanostructure is
simple, this procedure leads to a set of coupled equations of motion for
the steps. Our goal in the present work is to construct and solve these
equations of motion for the simple case of an infinite crystalline cone.
A partial account of this work is found in Ref.\ \onlinecite{israeli98}.
The
kinetic step model for the cone is derived in section II. In section
III we carry out numerical simulations of the model and examine the
evolution of surface morphology under various conditions.

Previous experimental and theoretical research on decay of
nanostructures has demonstrated that in most cases the surface reaches a
scaling state where the typical length scale depends on time
algebraically. Our simulations show that the cone profile also exhibits
such a scaling behavior. In section IV we show analytically that the
step flow model admits such solutions. We calculate the scaling
exponents and
derive a continuum equation for the scaling function. The properties of
the scaling function are analyzed in sections V and VI. 

It has been suggested that in some materials, steps may be permeable to
flow of atoms; i.e. atoms can hop between neighboring terraces without
attaching to the step which separates them. In section VII we discuss
the consequences of step permeability on the decay of an infinite cone,
and show that permeability does not change the qualitative behavior of
the system.   

\section{Step flow model of a crystalline cone}

We now consider a crystalline surface which 
consists of flat terraces, parallel to a high symmetry plane of the
crystal and separated by atomic steps. Islands and voids are ignored.
The evolution of such systems of steps was treated
long ago by Burton, Cabrera and Frank\cite{BCF} (BCF).
The BCF theory assumes that mass transfer between terraces is governed
by diffusion of adatoms on the terraces. These atoms are emitted
and absorbed at step edges, which according to BCF act as perfect sinks
and sources. This
last assumption, however, is valid only when adatom diffusion is the
rate limiting process. Modifications of the BCF model account for the
finite rate of adatom attachment-detachment processes at 
step edges. The BCF 
model was also generalized to include 
elastic and entropic interactions between
steps (see for example Refs.\
\onlinecite{OzdemirZangwill,BalesZangwill}). In
this section we construct such a generalized BCF model for the
morphological evolution of a conic hill on a crystal surface.      
    
\begin{figure}[h]
\centerline{
\epsfxsize=90mm
\epsffile{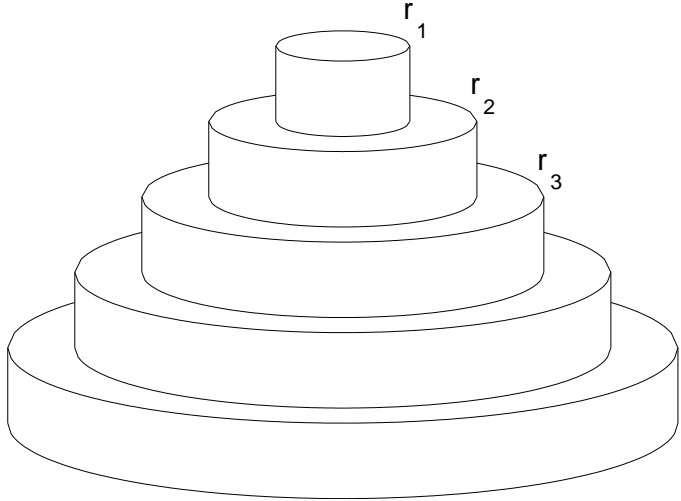}}
{\vbox{\vspace*{0.6cm}}}
\caption{Illustration of the step configuration in a crystalline cone.
The steps height is exaggerated.} 
\label{cone_picture}
\end{figure}

Consider the surface of an infinite crystalline cone, which consists of
circular concentric steps of radii $r_i(t)$, separated by flat
terraces (Fig.\ \ref{cone_picture}).
The index $i$ grows in the direction away from the center of the
cone. 
These steps may absorb or emit atoms, which then
diffuse across the neighboring terraces with a diffusion constant
$D_s$. Assuming no deposition of new material, no evaporation and no
transport through the bulk, the adatom concentration $C_i(\vec r)$
on the $i$th terrace satisfies the diffusion equation  
\[ D_s\nabla^2C_i(\vec{r})=\frac{\partial C_i(\vec{r})}{\partial t}\;,\]
where $\vec{r}$ is a two dimensional vector parallel to the high
symmetry
terraces. In most situations, the time scale associated with step motion
is much larger than the time scale of surface diffusion. One can
therefore
assume that the adatom diffusion field is always in a steady
state; i.e., for any step configuration, the diffusion field
reaches a steady state before the steps move significantly. Within
this quasi-static approximation, the r.h.s.\ of the diffusion
equation can be neglected. Using the radial symmetry of the cone, we can
write the static diffusion equation in polar coordinates as:
\begin{equation}
\frac{\partial^2  C_i(r)}{\partial r^2}+\frac{1}{r} \frac{\partial
  C_i(r)}{\partial r}=0\;.
\label{eq:diffusion}
\end{equation}
The general solution of this equation is
\begin{equation}
  \label{eq:diff_solution}
  C_i(r)=A_i \ln{r}+B_i\;.
\end{equation}
The coefficients $A_i$ and $B_i$ are determined by the boundary
conditions at the step edges. To define these conditions,
we assume that
the flux of atoms at the two step edges bounding the $i$th terrace is
determined by first order kinetics, 
characterized by an attachment-detachment rate coefficient $k$:
\begin{eqnarray}
D_s \left. \frac{\partial  C_i }{\partial r}
\right|_{r_i}&=&k\left(\left.C_i
  \right|_{r_i}-C^{eq}_i\right)\; \nonumber \\
-D_s \left. \frac{\partial C_i }{\partial r}
\right|_{r_{i+1}}&=&k\left(\left.C_i
  \right|_{r_{i+1}}-C^{eq}_{i+1}\right)\;, 
\label{eq:boundry}
\end{eqnarray}
where $C^{eq}_i$ is the equilibrium concentration of atoms on the
terrace adjacent to the $i$th step. Using these boundary conditions, we
find that the constants $A_i$ are given by
\begin{equation}
  \label{eq:ais}
  A_i=\frac{C^{eq}_i-C^{eq}_{i+1}}
  {\ln{r_i}-\ln{r_{i+1}}-\frac{D_s}{k}\left(\frac{1}{r_i}+\frac{1}{r_{i+
  1}}
    \right)} \;.  
\end{equation}
Employing mass conservation at the step we obtain the step velocity:
\begin{equation}
\frac{dr_i}{dt}=\Omega D_s
\left. \left(\frac{\partial C_i}{\partial r}-\frac{\partial
C_{i-1}}{\partial r}\right) \right|_{r_i}
=\Omega D_s\frac{A_i-A_{i-1}}{r_i}\;,
\label{eq:velocity}
\end{equation}
where $\Omega$ is the atomic area of the solid.

In order to complete the solution of the diffusion problem we have to
determine the equilibrium adatom concentration at step edges. This
concentration depends on the local step curvature and on the radii of
neighboring steps. According to the Gibbs-Thomson relation,
$C^{eq}_i$ is given by
\begin{equation}
  \label{eq:gibbsthomson}
  C^{eq}_i=\bar{C}^{eq} \exp{\frac{\mu_i}{T}} \approx \bar{C}^{eq}
  \left(1+
  \frac{\mu_i}{T}\right)\;,
\end{equation}
in units where the Boltzmann constant is equal to 1.
Here $\mu_i$ is the chemical potential associated with the addition of
an atom to the solid at the $i$th step, $T$ is the temperature and
$\bar{C}^{eq}$ is the adatom equilibrium
concentration at the edge of a straight isolated step. 
To evaluate the
step chemical potential, $\mu_i$, we take into account the step line
tension, $\Gamma$, and a repulsive interaction between nearest neighbor
steps. The magnitude of this interaction is inversely proportional to
the square of the
distance between 
the steps at large distances. Such a dependence is consistent with
entropic as well as
elastic\cite{MarchenkoParshin,AndreevKosevich} interactions between
straight steps. We follow 
Ref.\ \onlinecite{TanakaBartelt} and take the interaction energy 
between step $i+1$ and an atomic segment of the $i$th step to be
\begin{equation}
U(r_i,r_{i+1})= \frac{G \sqrt{\Omega} r_{i+1}}{\left( r_i+r_{i+1} 
\right)\left(r_{i+1}-r_i\right)^2}\;,
\end{equation}
where $G$ is the interaction strength.
The chemical potential of the $i$th step is then given by
\begin{equation}
\mu_i=\frac{\Omega \Gamma}{r_i} +\sqrt{\Omega}\frac{\partial 
\left[ U(r_i,r_{i+1})+U(r_i,r_{i-1}) \right] }{\partial r_i}\;.
\label{eq:full_mu}
\end{equation}
This equation applies also to the
chemical potential of the first step if we set $r_0=0$.

As we show below, in the long time limit
the distance between steps is small compared with the step radius. In
this limit we can
approximate the step chemical potential by
\begin{equation}
\mu_i = \frac{\Omega \Gamma}{r_i} + \Omega G \left(
  \frac{2r_{i+1}}{r_{i+1}+r_i} \cdot
  \frac{1}{\left(r_{i+1}-r_i\right)^3} -
  \frac{2r_{i-1}}{r_i+r_{i-1}} \cdot
  \frac{1}{\left(r_i-r_{i-1}\right)^3}
\right)\;.
\label{eq:simple_mu}
\end{equation}
This approximation simplifies the algebra considerably, and we verified
that it does not affect the analytical and numerical results described
below.

We are now ready to write down the step equations of motion using
Eqs.\ (\ref{eq:ais}), (\ref{eq:velocity}) and the Gibbs-Thomson relation
(\ref{eq:gibbsthomson}). It is convenient to use 
dimensionless radii, $\rho_i$, and dimensionless time $\tau$: 
\begin{eqnarray*}
\rho_i&=&\frac{T}{\Omega \Gamma} \cdot r_i\;,  \\  
\tau&=&D_s
\bar{C}^{eq}\Omega\cdot\left(\frac{T}{\Omega\Gamma}\right)^2\cdot
\left(1+\frac{D_sT}{k\Omega\Gamma}\right)^{-1}\cdot t\;.
\end{eqnarray*}
In terms of these variables, the equations of motion take the form
\begin{eqnarray}
\dot{\rho}_i&\equiv&\frac{d\rho_i}{d\tau}=\frac{a_i-a_{i-1}}{\rho_i}\;,
\;\;
\mbox{with}  \label{eq:dimensionlessvelocity} \\ 
a_i&=&\frac{\xi_i-\xi_{i+1}}
{\left(1-q\right)\ln \frac{\rho_i}{\rho_{i+1}} - 
q\left(\frac{1}{\rho_i} + \frac{1}{\rho_{i+1}}\right)} \;, \nonumber \\
\xi_i&=& \frac{1}{\rho_i} + g \left(
  \frac{2\rho_{i+1}}{\rho_{i+1}+\rho_i} \cdot
  \frac{1}{\left(\rho_{i+1}-\rho_i\right)^3} - 
  \frac{2\rho_{i-1}}{\rho_i+\rho_{i-1}} \cdot
  \frac{1}{\left(\rho_i-\rho_{i-1}\right)^3} 
\right) \;.\nonumber  
\end{eqnarray}
Eqs.\ (\ref{eq:dimensionlessvelocity}) depend on two
parameters: $g=\frac{T^2G}{\Omega^2\Gamma^3}$
measures the strength of step-step interactions $G$ relative to
the line tension $\Gamma$,  
while the parameter $q=(1+\frac{k\Omega\Gamma}{D_sT})^{-1}$ determines
the rate limiting process in the system. When $q\rightarrow 1$,
diffusion across terraces is fast and 
the rate limiting process is attachment and
detachment of adatoms to and from steps. On the other hand, when
$q\rightarrow 0$, 
the steps act as perfect sinks and the rate limiting process is 
diffusion across terraces. 

\section{Results of simulations}

We integrated Eqs.\ (\ref{eq:dimensionlessvelocity}) numerically 
both in the case of diffusion limited kinetics (DL)
and in the case of attachment-detachment limited kinetics (ADL). The
initial configuration was a uniform step train. In principle,
the initial step separation is a parameter of the model. However, for
any value of this parameter one can change the units of length and time
and get the same equations of motion with initial step separation of
unity and different values of the parameters $g$ and $q$. Thus it is
sufficient
to consider an initial step separation of unity.

\begin{figure}[h]
\centerline{
\epsfxsize=80mm
\epsffile{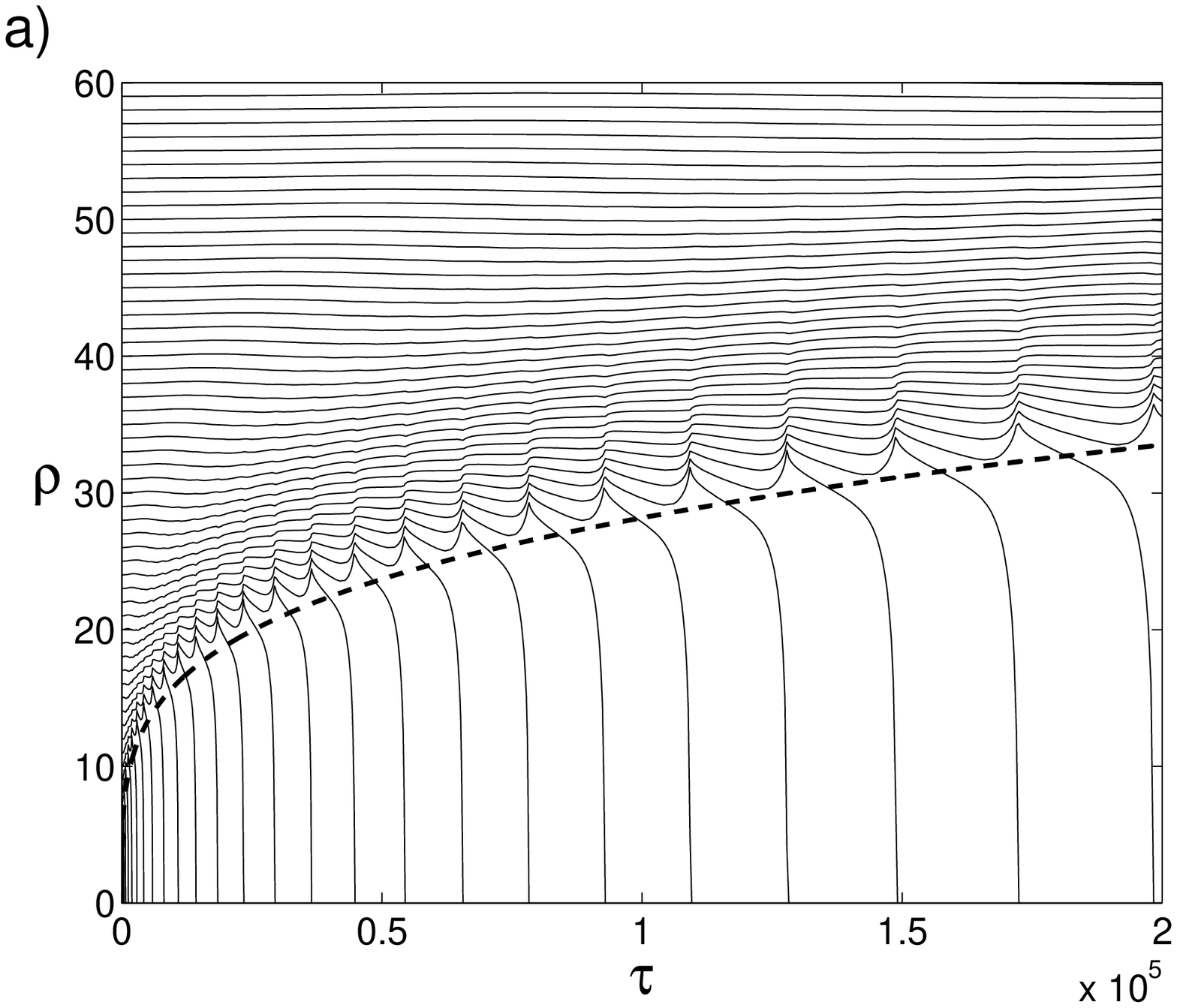}  
\hspace{0.1in}
\epsfxsize=80mm
\epsffile{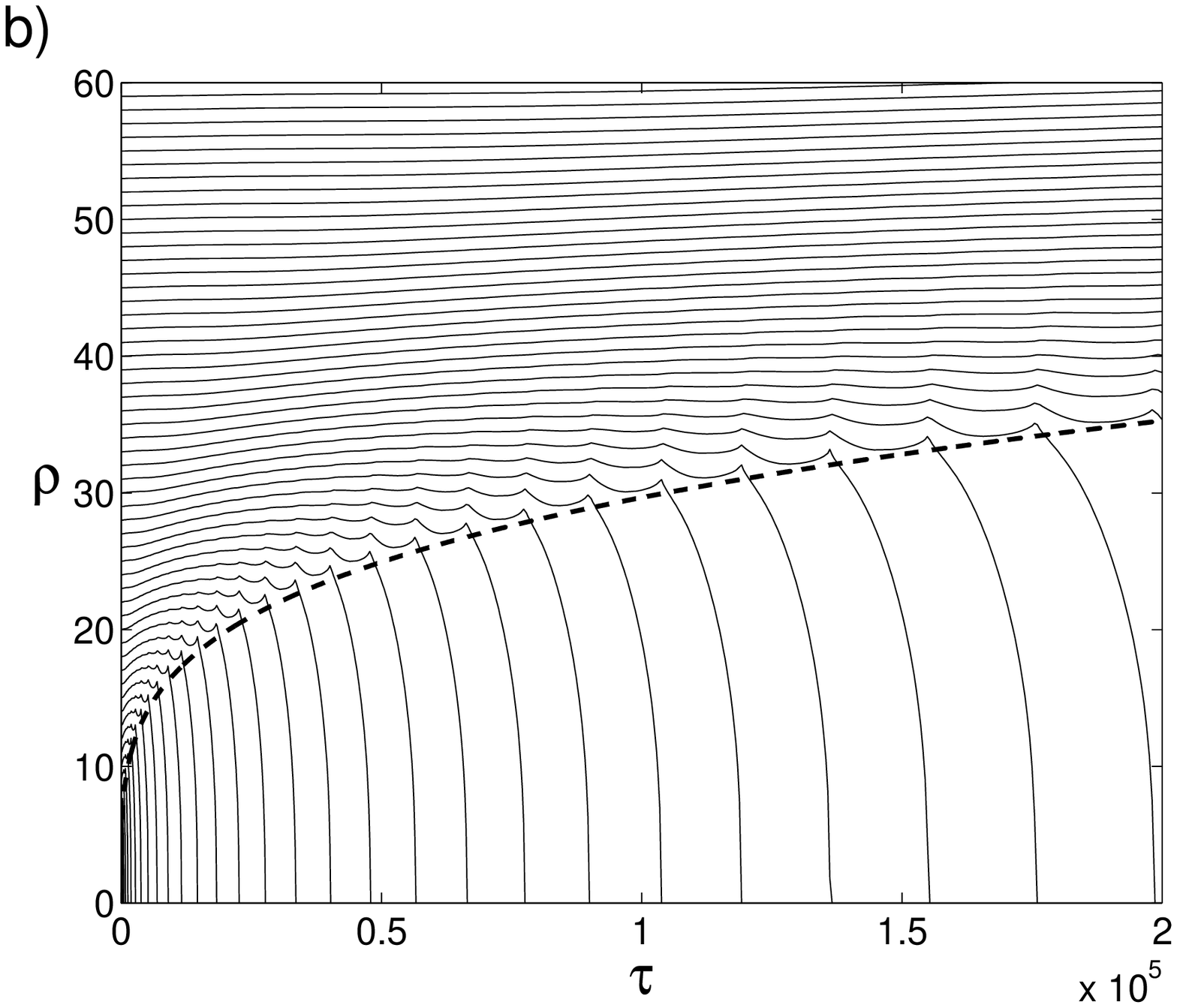}}
\vspace{0.1in}
\caption{Time evolution of the step radii in the (a) ADL and (b) DL  
  cases with $g=0.01$. The radius of the facet edge can be fitted by a
  $\tau^{1/4}$ power
  law (dashed lines).} 
\label{traces}
\end{figure}

When the repulsive interactions 
between steps are weak (i.e. $g$ is small), there is a striking
difference
between the dynamics in the DL and ADL limits. In the ADL
case the system becomes unstable towards step bunching, whereas in the
DL case there is no such instability. However, when $g$ is large
enough the instability disappears even in the ADL
case. Let us first discuss situations where the step bunching
instability
does not occur. Fig.\ \ref{traces}
shows the time evolution of the system in the ADL and the DL cases
with a relatively large value of $g$.
Each line in the figures describes the evolution of the radius of one
step. We note that
the innermost step shrinks while the other steps expand and absorb the
atoms
emitted by the first step. When the innermost step disappears, the next
step starts shrinking and so on. Our observations indicate that the time
at which the $n$th
step disappears, $\tau_n$, grows with $n$ as $\tau_n\sim n^4$.
This process results in a propagating front, which leaves a
growing plateau or a facet at the center of the cone. At large times,
the (dimensionless) position of this front behaves as
$\rho_{\text{front}}(\tau)\sim \tau^{1/4}$. This is shown by the
dashed lines in Fig.\ \ref{traces}. 

This power law is an indication of a much more
general and interesting phenomenon. It turns out that at large times, 
not only the front position but also the positions of
minimal and maximal step densities scale as $\tau^{1/4}$. In
fact, the step density $D(\rho,\tau)$, defined as the inverse step
separation, obeys the following scaling scenario:
There exist scaling exponents $\alpha$, $\beta$ and $\gamma$, which
define the scaled 
variables $x\equiv \rho\tau^{-\beta/\gamma}$
and $\theta\equiv\tau^{1/\gamma}$. In terms of these variables
\begin{equation}
  \label{eq:scaling}
D(\rho,\tau)=\theta^\alpha F(x,\theta)\;,
\end{equation}
where the scaling function
$F$ is a {\em periodic} function of $\theta$
with some period $\theta_0$. 
Our ansatz is somewhat weaker than standard
scaling hypotheses, which would assume the scaling function $F$
is independent of $\theta$.
We introduce this periodic dependence because our simulations strongly
indicate a periodic behavior, generated by the motion of the
first step (see Fig.\ \ref{traces}). Thus the
disappearance time of step $n$ is $\tau_n=(n\theta_0)^\gamma$.
An immediate consequence of the scaling ansatz is that
if we define $\theta=\bar\theta+n\theta_0$ with
$0\leq\bar\theta<\theta_0$,
and plot
$\theta^{-\alpha}
D\left(\rho,\tau\right)$
against $x$, all the data with different values
of $n$ and the {\em same} value of $\bar\theta$ collapse onto a single
curve, $F(x,\bar{\theta})$.

\begin{figure}[h]
\centerline{
\epsfxsize=80mm
\epsffile{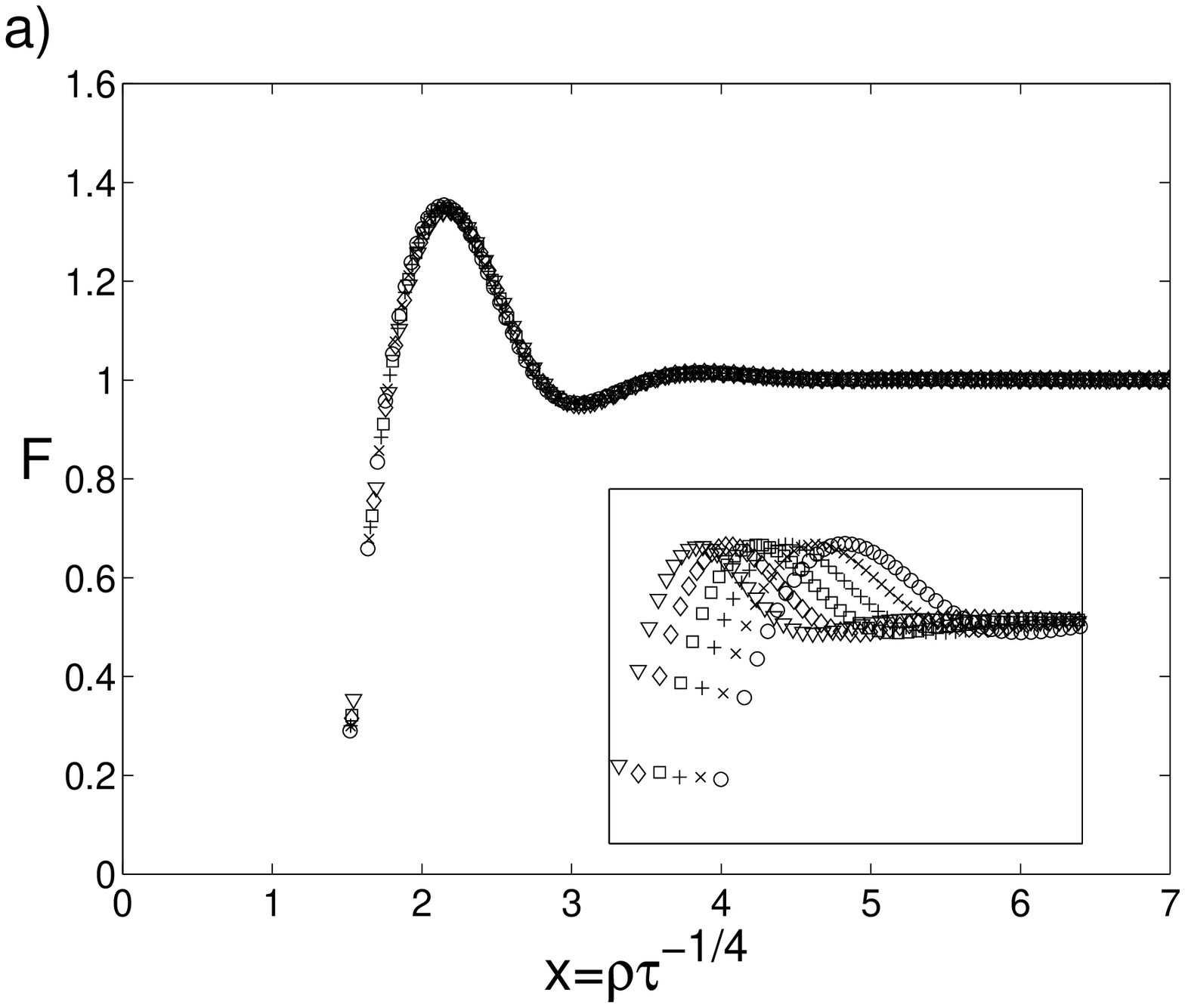}  
\hspace{0.1in}
\epsfxsize=80mm
\epsffile{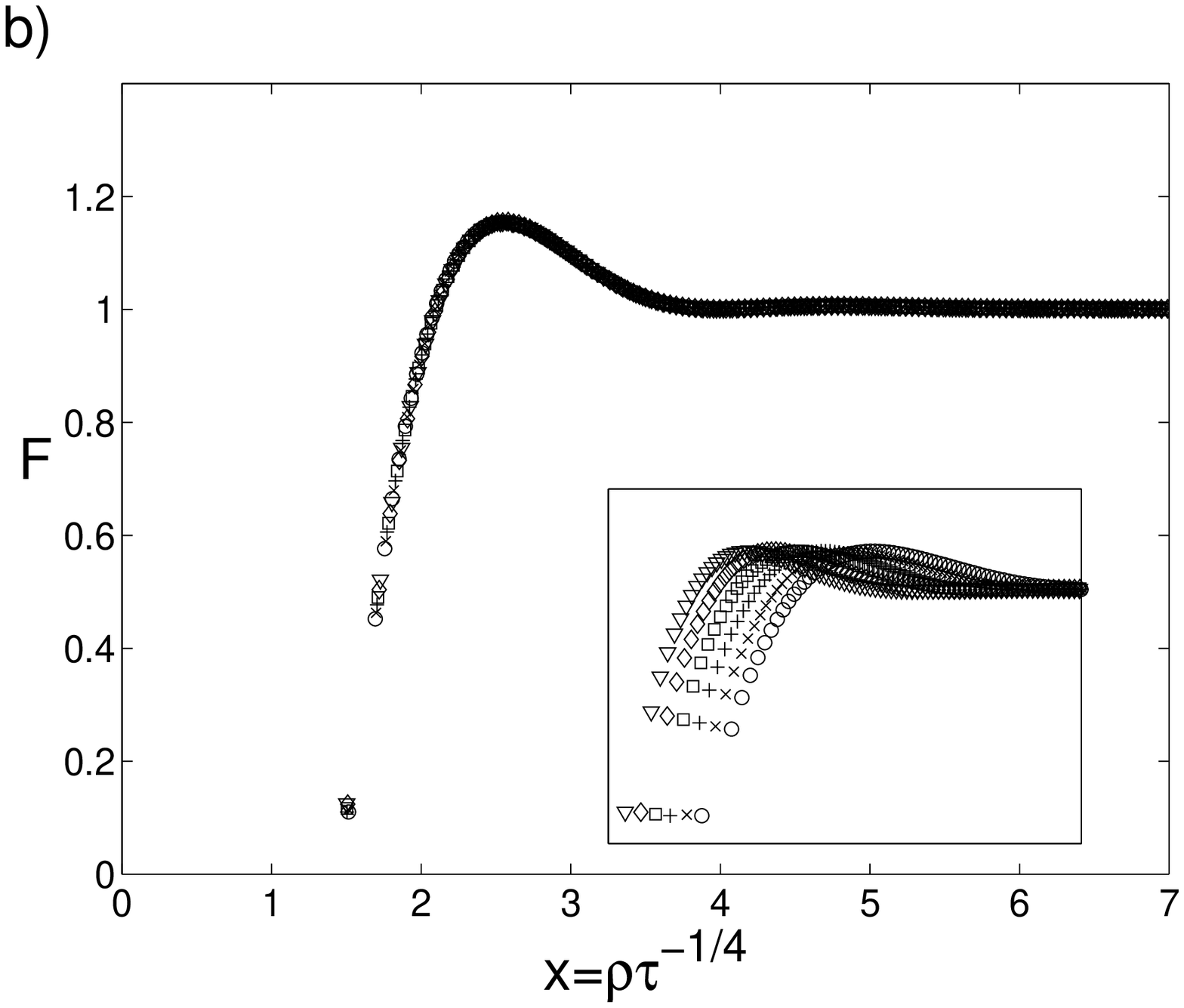}}
\vspace{0.1in} 
\caption{Data collapse of the density function in the (a) ADL and (b) DL
  cases with $g=0.01$. The values of the scaling exponents
  used here are $\alpha=0$, $\beta=1$ and $\gamma=4$. These figures
  show density functions with 6 
  different values of $n$ and the same value of $\bar\theta$, as a
  function
  of $x=\rho\tau^{-1/4}$. Different symbols corresponds to different
  values of $n$.
  The unscaled data is shown in the insets.} 
\label{scaling}
\end{figure}

To verify that our system obeys this scaling ansatz, we define the step
density at a discrete set of points in the middle of the terraces:
\begin{equation}
D\left(\frac{\rho_i(\tau)+\rho_{i+1}(\tau)}{2},\tau\right)=\frac{1}{\rho
_{i+1}
(\tau)-\rho_i(\tau)}\;.
\label{eq:discretedens}
\end{equation}
Fig.\ \ref{scaling} shows plots of
$D(\rho,\tau)$ as a function of   
$x=\rho\tau^{-1/4}$ for a fixed value of $\bar\theta$
and 6 different values of $n$
in the ADL and the DL cases.
The excellent data collapse shows that our scaling ansatz indeed holds
with $\alpha=0$, $\beta=1$ and $\gamma=4$ in both cases.

We now turn to discuss effects of interactions between steps. As
shown above, the behavior of the cone in the ADL and the DL limits is
very similar when the repulsion between steps is strong. We have already
mentioned that when the repulsion is weak, the behavior of the system in
the DL limit is very different from its behavior in the ADL case. This
is not surprising,
since linear stability analysis (see Appendix A) of the two cases in
the absence of 
step-step interactions predicts that the ADL case is unstable
towards step bunching while the DL case is marginal. In intermediate
cases ($0<q<1$) the system is unstable when the
interactions are sufficiently weak. 

\begin{figure}[h]
\centerline{
\epsfxsize=80mm
\epsffile{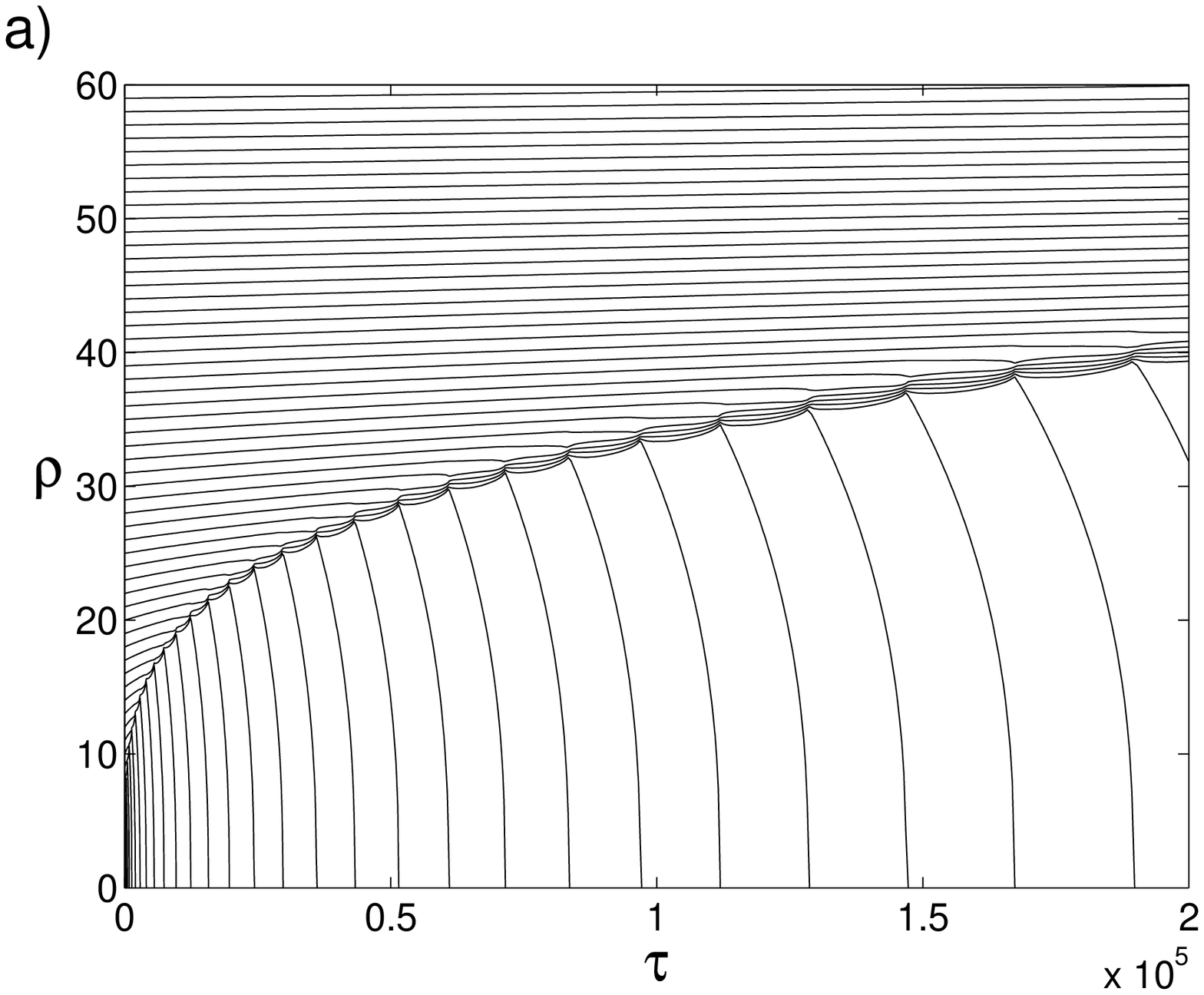}
\hspace{0.1in}
\epsfxsize=80mm
\epsffile{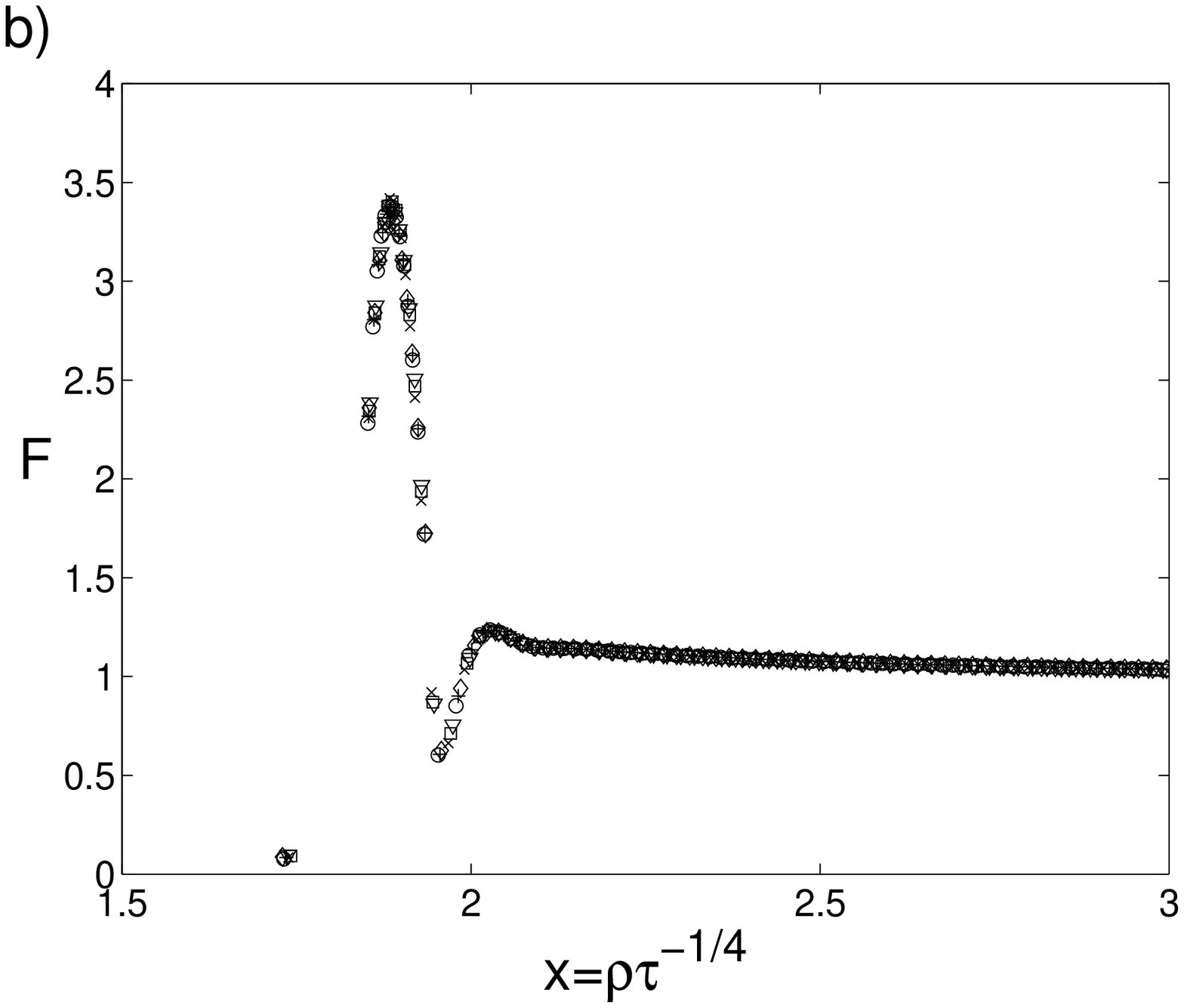}}  
\vspace{0.1in}
\caption{(a) Time evolution of the step radii in the DL
  case with $g=10^{-6}$. (b) Data collapse of the density function in
  the same system. Different symbols correspond to different
  times.}
\label{DL_lowint}
\end{figure}

Fig.\
\ref{DL_lowint} shows the time evolution and the data
collapse of the 
density function in the DL case when step-step interactions are
weak. As one can see, the behavior of the cone is qualitatively similar
to the large $g$ example (Fig.\ \ref{traces} (b)) and that the
scaling ansatz still holds. Quantitatively, the step density near the
facet edge is much higher for small values of $g$. Also the dependence
of the scaling function on scaled time within a collapse period is much
more
pronounced when the interactions are weak. 

\begin{figure}[h]
\centerline{
\epsfxsize=80mm
\epsffile{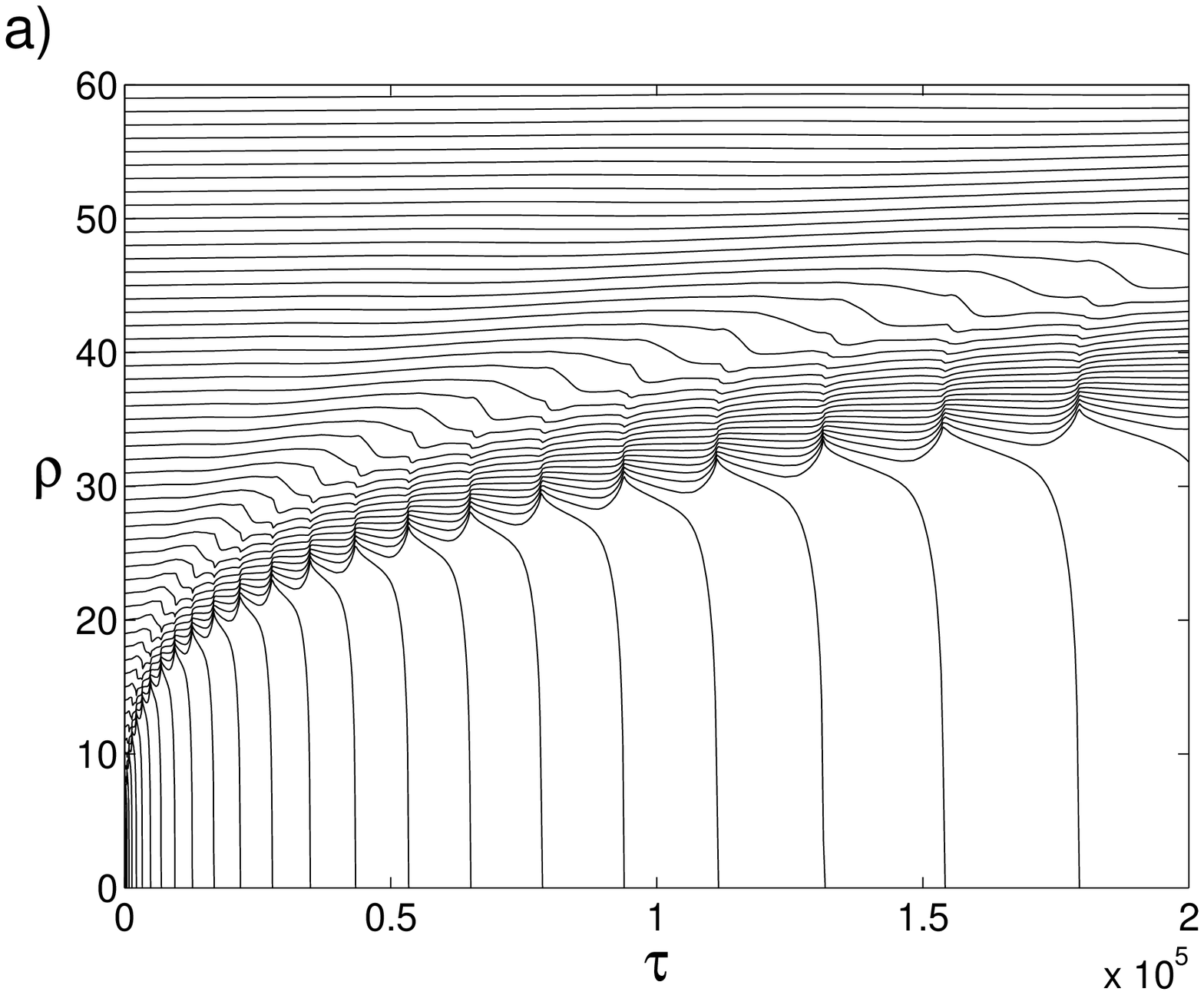}  
\hspace{0.1in}
\epsfxsize=80mm
\epsffile{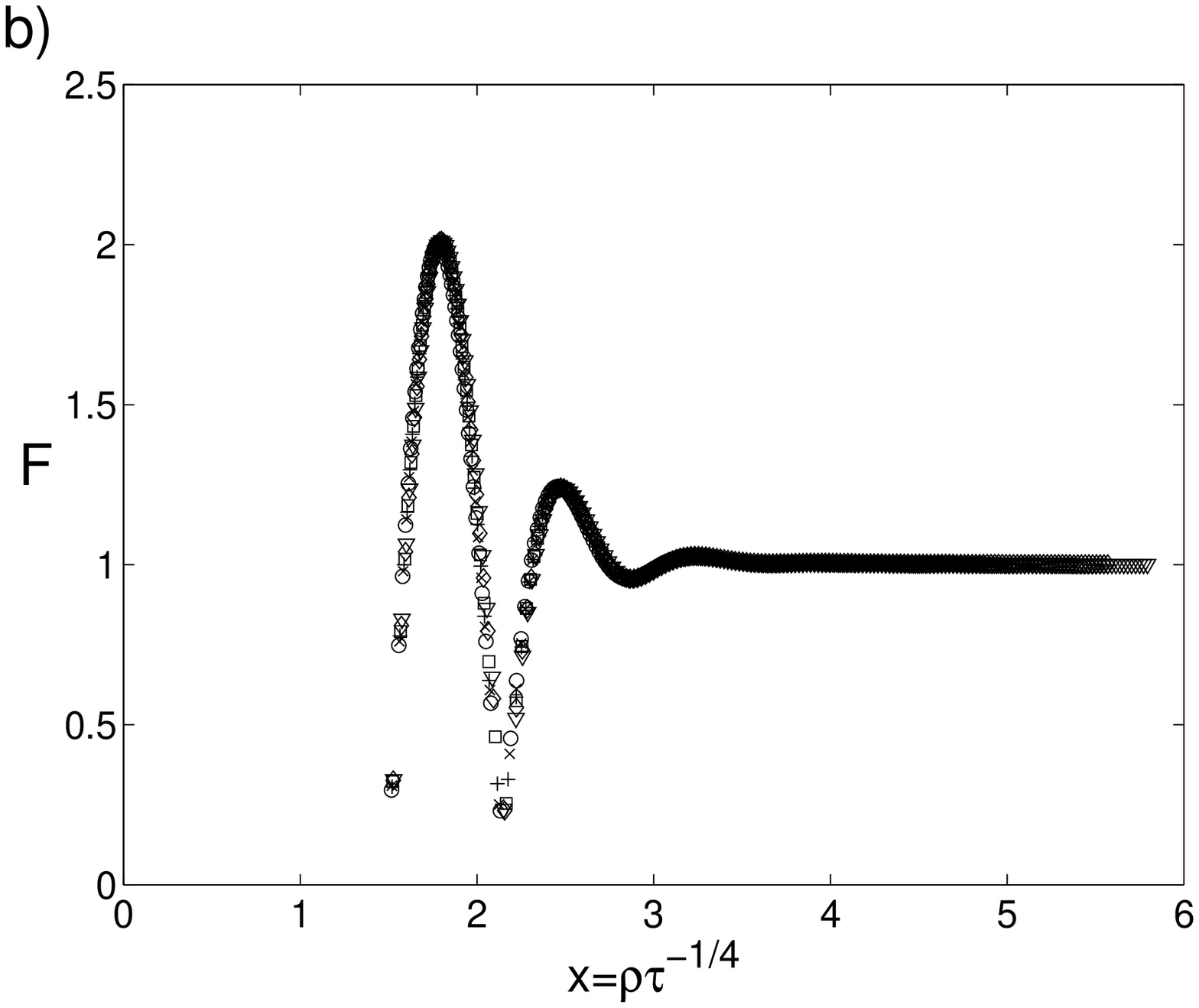}}
\centerline{
\epsfxsize=80mm
\epsffile{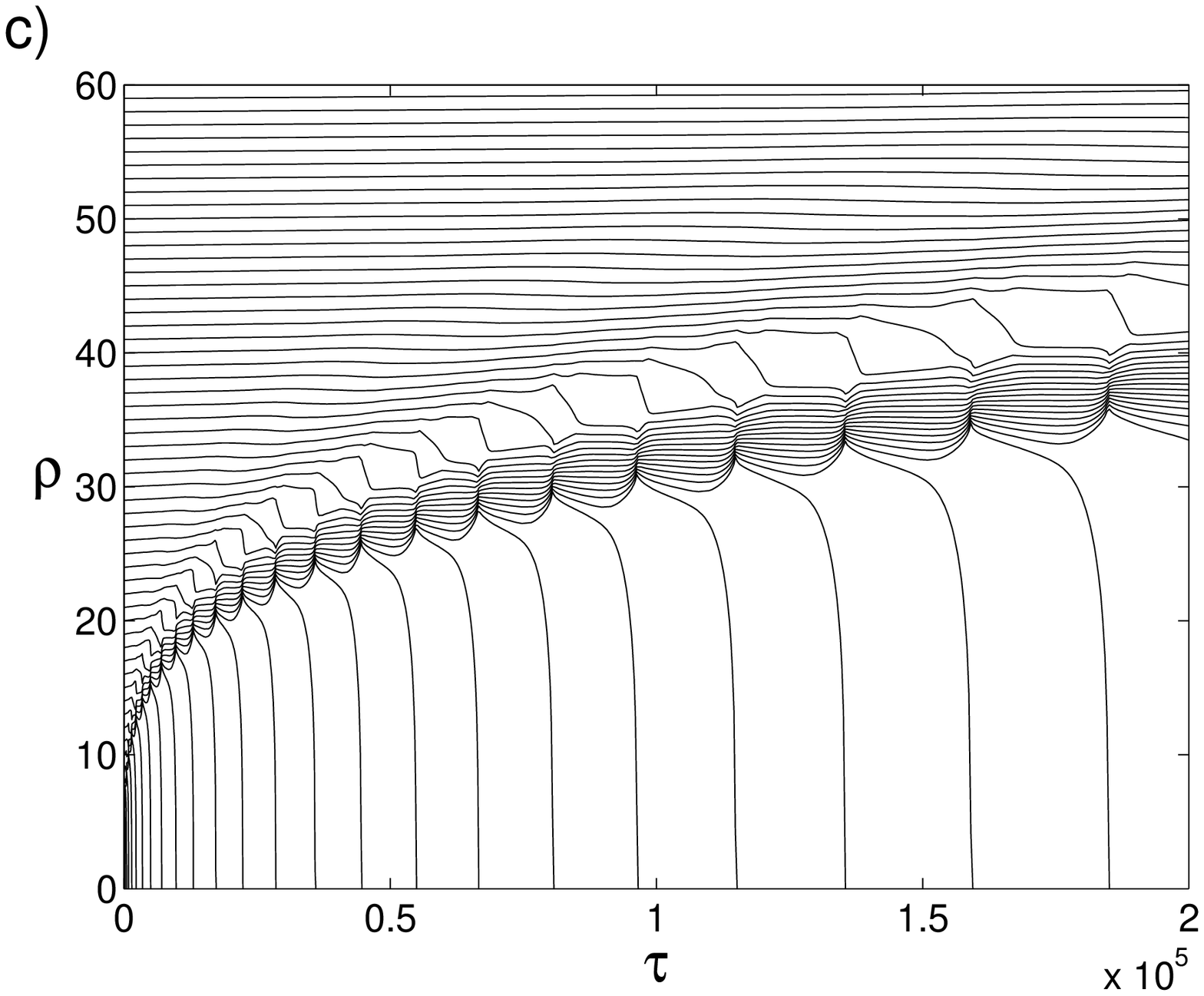}  
\hspace{0.1in}
\epsfxsize=80mm
\epsffile{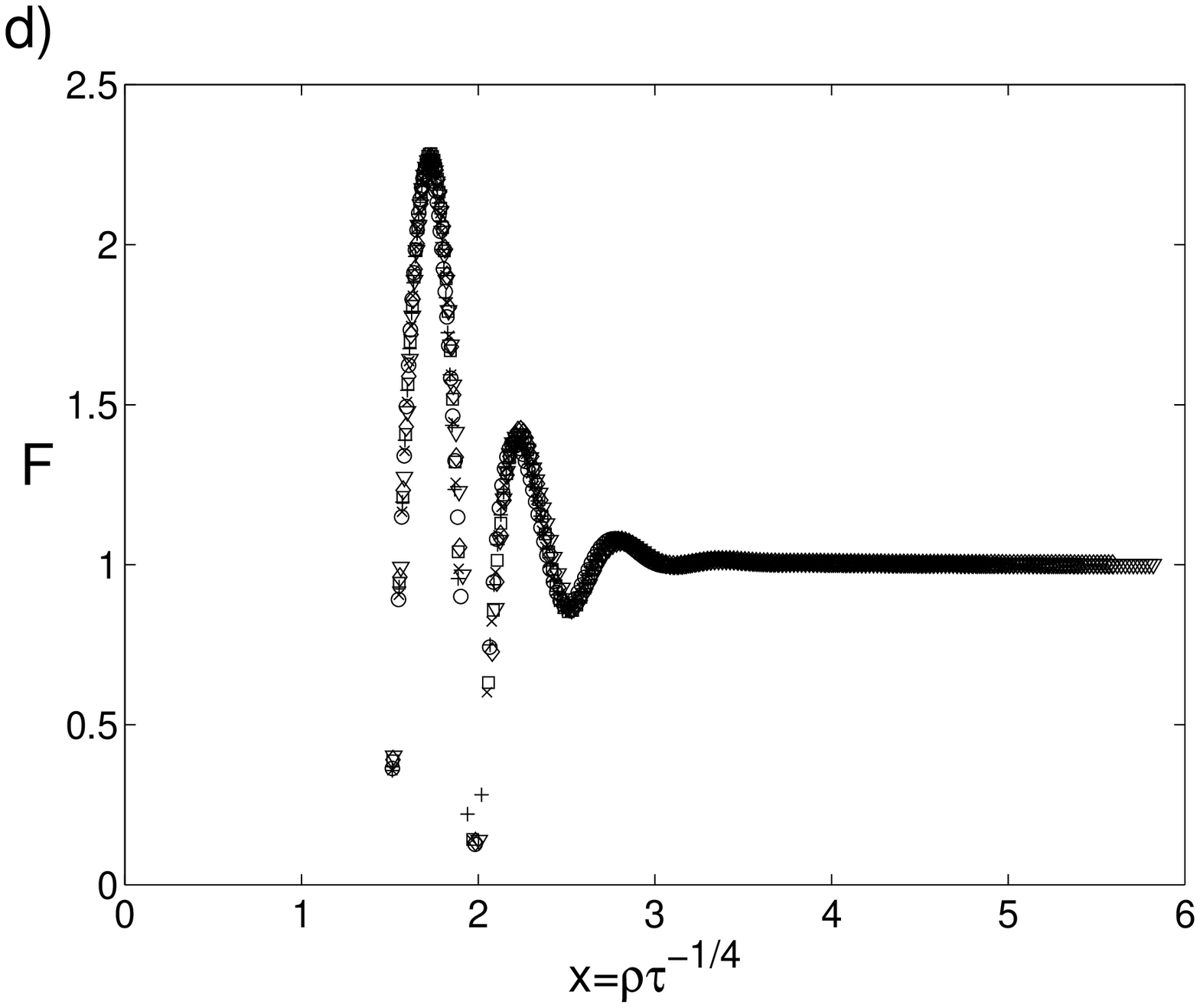}}
\centerline{
\epsfxsize=80mm
\epsffile{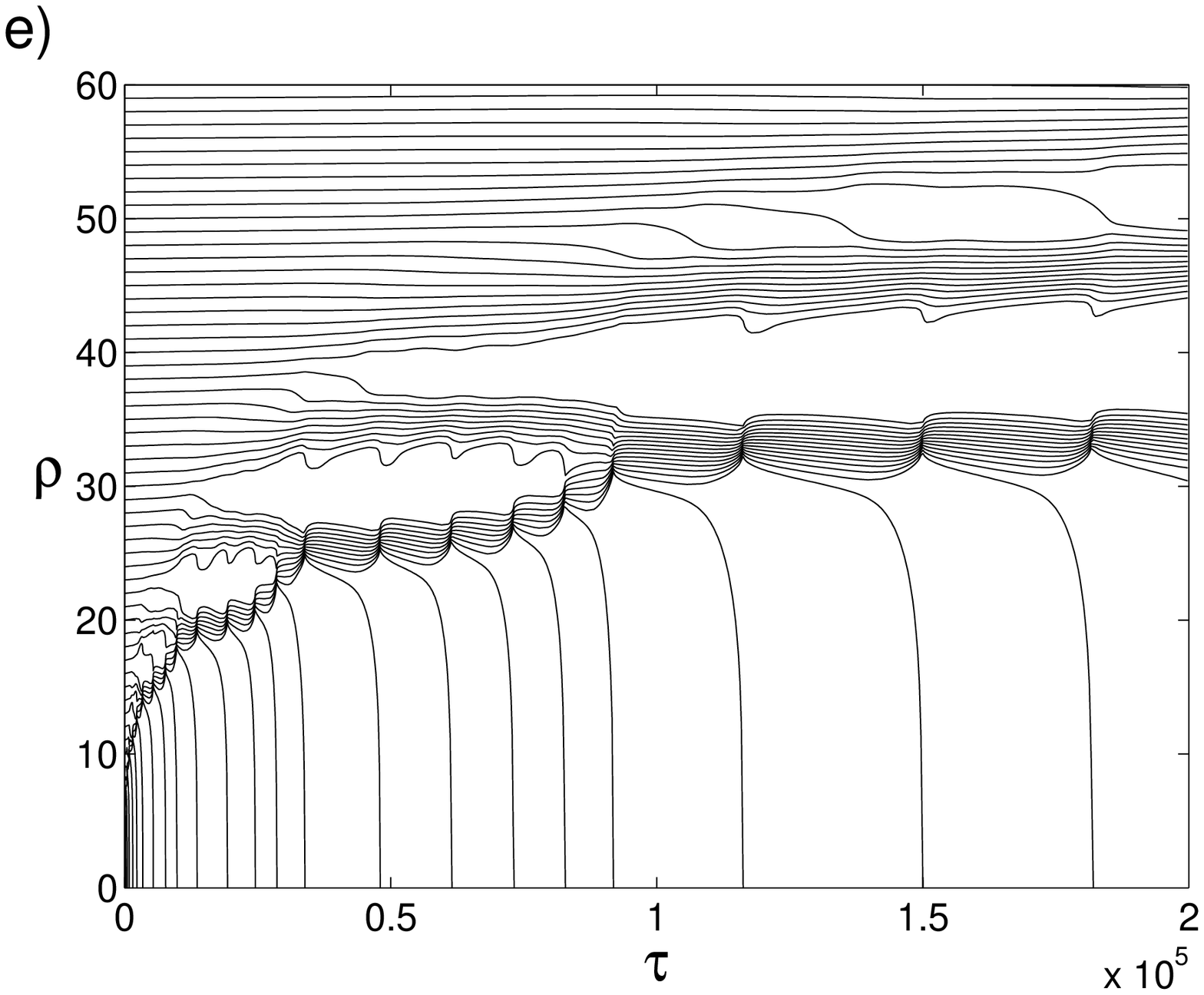}  
\hspace{0.1in}
\epsfxsize=80mm
\epsffile{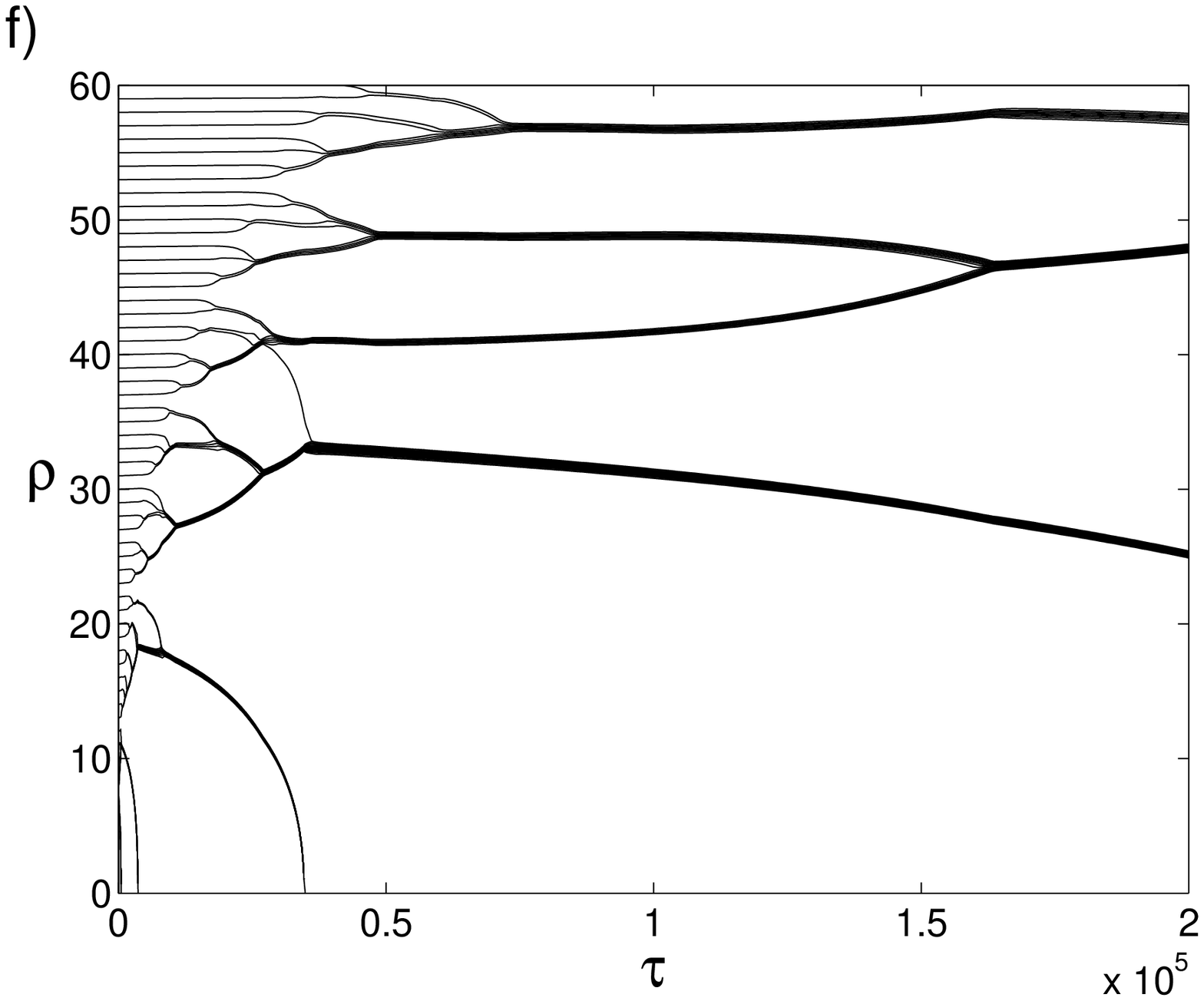}} 
\vspace{0.1in} 
\caption{Time evolution and scaled density functions of ADL systems
  with different interaction strengths. (a) and (b): $g=10^{-3}$, 
  (c) and (d): $g=5\cdot 10^{-4}$, (e): $g=2.5\cdot 10^{-4}$ and (f):
  $g=10^{-6}$.} 
\label{transition}
\end{figure}

The dependence of step kinetics on the strength of the interactions is
much more complicated in the ADL case. Fig.\
\ref{transition} shows the evolution and when possible, the scaled
density function of a 
series of ADL systems which differ in the value of $g$. The two
quantitative observations we made in the DL case 
hold also here: The high step density region near the front
becomes more dense, and the time dependence of the scaling
function becomes more pronounced as the value of $g$ is reduced (Fig.\
\ref{transition} (a)-(d)). In addition, we noted that the approach of
the
system towards the scaling state is slower. Adjacent to the dense
region at the edge of the facet is another region of very low
density of steps. This region becomes less dense as the value of $g$ is
reduced. Below a critical value, $g_c$, of the interaction parameter,
a few steps between the low density region and the facet edge form 
a bunch. At this stage the scaling ansatz breaks down.
The system no longer exhibits a simple periodic nature, which results
from the
collapse of single steps, one at a time. Instead, it seems to adopt a
complicated almost 
periodic pattern which involves the collapse of many steps in each 
period (Fig.\ \ref{transition} (e)). Hints for this periodicity
changes are already present in Fig.\ \ref{transition} (a) and (c),
where the steps crossing the low density region follow a threefold
periodicity.
If the value of $g$ is further reduced, bunches of
steps collapse 
together, and finally the whole system
becomes unstable toward step bunching (Fig.\ \ref{transition}
(f)). Neighboring steps  
merge and form a bunch, which in turn merges with another bunch and so
on. Step bunches rather than isolated steps become the dynamic objects
in the system. 

\section{Scaling analysis and the continuum model}

The above results suggest that when the step bunching
instability does not occur, the time evolution of the system can be
described by a step density function, which is continuous in
both position and time variables.
In this section we derive such a continuum model by carrying out
a scaling analysis. We obtain an equation which governs the scaling
function and evaluate the scaling exponents analytically.

Motivated by the
simulation results, we assume that at long times the scaling ansatz,
Eq.\ (\ref{eq:scaling}), holds.
This together with conservation of the total volume of the system,
already determines the two scaling exponents
$\alpha$ and $\beta$. First, we derive a relation between these 
scaling exponents
by considering the height profile $h(\rho,\tau)$. 
Assuming steps of unit
height, the height difference between two points on the surface is
given by the number of steps between them. The continuous
analog of this statement can be used to derive the following relation
between the height profile and the step density: 
\begin{equation}
h(\rho,\tau)=h_0(\tau)-\int_{0}^{\rho}D(\rho',\tau)d\rho'\;,
\label{eq:height}
\end{equation} 
where $h_0(\tau)$ is the height at the origin.
Far enough (in the limit $\rho\rightarrow\infty$), $h(\rho,\tau)$ does
not change
with time. Calculating the height change at infinity using Eq.\
(\ref{eq:height}) we find
\begin{equation}
  \label{eq:hzero}
  h_0(\tau)-h_0(0)-
  \tau^{\frac{\alpha+\beta}{\gamma}}
\int_{0}^{\infty}\left[F(x,\theta)-F(\infty,0)\right]dx=0\;,
\end{equation}
where we have changed the integration variable and used the definition
(Eq.\ (\ref{eq:scaling})) of the
scaling function $F$. 
On the other hand, $h_0(0)-h_0(\tau_n)=n$ because $\tau_n$ is the time
of disappearance of the $n$th step. $\gamma$ satisfies the relation
$\tau_n\sim n^\gamma$, and therefore we have
$h_0(0)-h_0(\tau_n)\sim \tau_n^{1/\gamma}$.
This and the $\tau^{(\alpha+\beta)/\gamma}$ dependence in Eq.\
(\ref{eq:hzero}) lead to the relation $\alpha+\beta=1$. 
 
In addition, conservation of the total volume of the system, ${\cal V}$,
implies that 
\begin{equation}
  \label{eq:volume}
{\cal V}=2\pi\int_0^\infty \rho h(\rho,\tau)d\rho
\end{equation}
is independent of $\tau$.
Integration by parts of the derivative of this integral with respect
to $\tau$ yields the following equation:
\begin{equation}
\int_0^\infty \rho^2\frac{\partial D(\rho,\tau)}{\partial \tau}
d\rho=0\;.
\label{eq:conservation}
\end{equation}
Evaluating this integral in terms of the function $F$ and
the scaled variables $x$ and $\theta$ we obtain the equation
\begin{equation}
\label{eq:conservation2}
\int_{0}^{\infty}x^2
\left[\frac{\alpha F(x,\theta)}{\theta}+\frac{\partial
    F(x,\theta)}{\partial \theta} \right]dx=0\;.
\end{equation}
This can be satisfied for all $\theta$ only if $\alpha=0$, since $F$ is
a periodic function of $\theta$. Combining this result with the
previously obtained relation $\alpha+\beta=1$,
we conclude that
\begin{equation}
  \label{eq:alphabeta}
  \alpha=0\; \;, \; \beta=1\;.
\end{equation}

Thus, $\gamma$ is the only nontrivial scaling exponent in the model.
To evaluate $\gamma$ and the scaling function $F$,
we continue with
the equation for the full time derivative of the step
density $D$:
\begin{equation}
\frac{dD}{d\tau}=\frac{\partial D}{\partial \tau}+\frac{\partial
D}{\partial
  \rho} \cdot\frac{d \rho}{d\tau}\;.   
\label{eq:partials}
\end{equation}
Eq.\ (\ref{eq:partials}) can be evaluated in the middle of the terrace
between
two steps (i.e. at $\rho=(\rho_i+\rho_{i+1})/2$). The l.h.s.\
of (\ref{eq:partials}) is calculated by taking the time derivative of
Eq.\ (\ref{eq:discretedens}): $dD/d\tau=-D^2
(\dot{\rho}_{i+1}-\dot{\rho}_i)$.
This together with the fact that
$d \rho/d \tau=(\dot\rho_i+\dot\rho_{i+1})/2$
leads to the relation 
\begin{equation}
\frac{\partial D}{\partial
        \rho}\frac{\dot{\rho}_{i+1}+\dot{\rho}_i}{2} 
        +\frac{\partial D}{\partial \tau} 
        +D^2(\dot{\rho}_{i+1}-\dot{\rho}_i)=0\;. 
\protect \label{eq:partials2}
\end{equation}

Now we change variables to $\theta$ and $x_i\equiv\rho_i\theta^{-1}$
(since $\beta=1$),
and transform Eq.\ (\ref{eq:partials2}) into an equation for the scaling 
function $F$:
\begin{equation}
  \label{eq:partial3}
  \frac{\partial F}{\partial x}\left(\theta^{\gamma-1}
    \frac{\dot{\rho}_{i+1}+\dot{\rho}_i}{2}-\frac{x}{\gamma}\right)
  +\frac{\theta}{\gamma}\frac{\partial F}{\partial \theta}
  +F^2 \theta^\gamma \left(\dot{\rho}_{i+1}-\dot{\rho}_i\right)=0\;.
\end{equation}
The step velocities $\dot\rho_i$ and $\dot\rho_{i+1}$ can also be
expressed in terms of the $x_i$'s, but we defer this algebraic
manipulation to a later stage. 

Our next goal is to take a continuum limit of Eq.\ (\ref{eq:partial3})
in
the variable $x=(x_{i+1}+x_i)/2$. Such a continuum limit becomes exact
in the long time limit. To see this,
let us rewrite Eq.\ (\ref{eq:discretedens}) in terms of $x_i$'s,
$\theta$ and $F$:
\begin{equation}
x_{i+1}-x_{i}=\frac{\theta^{-1}}{F\left((x_{i+1}+x_i)/2,\theta
\right)}\;.
\label{eq:xsdifference}
\end{equation}
According to this equation, the difference between successive $x_i$'s is  
of order $\theta^{-1}$ wherever $F$ does not vanish. 
In the large $\theta$ (long time) limit
these differences become vanishingly small. This justifies the continuum
limit in the variable $x$.

In practice, we take the continuum limit in the following way. 
We evaluate the function $F$ at the position
$(x_{i+k}+x_{i+k+1})/2$ by using it's Taylor expansion
\begin{equation}
  \label{eq:taylorexpansion}
  F\left(\frac{x_{i+k}+x_{i+k+1}}{2},\theta\right)\equiv
  \frac{\theta^{-1}}{x_{i+k+1}-x_{i+k}}=
  \sum_{n=0}^{\infty}\frac{1}{n!}\frac{\partial^n
  F(x,\theta)}{\partial
  x^n}\left(\frac{x_{i+k}+x_{i+k+1}}{2}-x\right)^n\;.
\end{equation}
As long as $k$ is finite, the difference
$x_{i+k}-x$ is small in the long time limit. It is therefore useful to
expand $x_{i+k}$ around $x$
\begin{equation}
  \label{eq:xisexpansion}
  x_{i+k}=x+\sum_{n=1}^{\infty}\phi_{kn}\theta^{-n}\;,
\end{equation}
and insert this expansion into Eq.\ (\ref{eq:taylorexpansion}). The
resulting equation is expanded as a power series in $\theta^{-1}$. By
requiring that the equation be obeyed to all orders in $\theta^{-1}$, we
can calculate the coefficients $\phi_{kn}$ for any desired $k$ and
$n$. These coefficients will involve the function $F$ and its
derivatives with respect to $x$, which are all periodic functions of
$\theta$. 

Next, we express the velocities $\dot{\rho}_i$ and
$\dot{\rho}_{i+1}$ in terms of the scaled radii $x_{i-2}$, $x_{i-1}$,
\ldots, $x_{i+3}$ using Eq.\ (\ref{eq:dimensionlessvelocity}) and the
transformation to scaled variables. We then use Eq.\
(\ref{eq:xisexpansion}) to expand the velocities in powers of
$\theta^{-1}$. The result of this expansion is
\begin{eqnarray}
  \label{eq:expandvelocities}
  \dot{\rho}_{i+1}+\dot{\rho}_i&=&\theta^{-3}({\cal
    A}+O(\theta^{-1}))\;, \nonumber \\
  \dot{\rho}_{i+1}-\dot{\rho}_i&=&\theta^{-4}({\cal
    B}+O(\theta^{-1}))\;.
\end{eqnarray}
${\cal A}$ and ${\cal B}$ are known expressions involving
$F$, $F'$, $F''$, $F'''$, $F''''$, where the primes denote partial
derivatives with respect to $x$. The full expressions are given in
Appendix B. The existence of derivatives up to
fourth order in this equation is a consequence of the fact that each
step ``interacts'' with four other steps (two on each side) through
the equations of motion (\ref{eq:dimensionlessvelocity}). Inserting
Eqs.\ (\ref{eq:expandvelocities}) into Eq.\ (\ref{eq:partial3}), we
obtain the following differential equation for F:
\begin{equation}
  -F'\cdot \frac{x}{\gamma}+
  \theta^{\gamma-4}\left(F'\cdot \frac{{\cal A}}{2} +
    F^2{\cal B} \right) 
  +\frac{\theta}{\gamma}\cdot \frac{\partial F}{\partial
    \theta}+O(\theta^{\gamma-5})=0\;.
  \label{eq:partials4}
\end{equation} 

Consider Eq.\ (\ref{eq:partials4}) at large $\theta$. Our expansion in
the small parameter $\theta^{-1}$ is valid only at values of
$x$ where $F$ does not diverge or vanish (see above). 
Therefore, the first term in
Eq.\ (\ref{eq:partials4}) is $O(1)$. This term has to be canceled by
the second term if we require $F$ to satisfy a single differential
equation. Hence, we must have 
\begin{equation}
\gamma=4\;.  \label{eq:gammarelation}
\end{equation}
The fourth term vanishes as $\theta\rightarrow\infty$
since $\gamma-5<0$, and the
third term must vanish as well. Therefore, in the large $\theta$ limit,
$F$ is only a function of $x$, and we are left with an
ordinary differential equation for $F$: 
\begin{equation}
F'\left(\frac{{\cal A}}{2}-\frac{x}{4}\right)
+F^2{\cal B}=0\;. 
\label{eq:differentialeq}
\end{equation}
The detailed form of this equation in the DL and ADL cases is
also given in Appendix B.

Let us emphasize several important properties of our scaling analysis.
First,
the values of the scaling exponents we calculated ($\alpha=0$, $\beta=1$
and $\gamma=4$) are consistent with the results of numerical simulations
(see above). Secondly, our continuum model is valid for
arbitrarily large surface curvature and slope (unlike other treatments
\cite{OzdemirZangwill,RettoriVillain}). Moreover, since our model is
an expansion 
in the truly small parameter $\theta^{-1}$ (see Eq.\
(\ref{eq:xsdifference})) it becomes {\em exact} in the large $\theta$
(long time) limit. Finally, 
note that in going to the continuum limit we lost the periodic
dependence
of $F$ on $\theta$. This periodicity is generated by the first step
which follows a unique equation of motion. We did not incorporate
this unique behavior into the continuum model and therefore should not
be surprised that this information is lost.

As we emphasized, our continuum model is an exact representation of
the original discrete  
system. For this reason it is interesting to compare it with other
continuum models, which do  
not emerge as limits of discrete systems of steps. Many authors use the 
continuity equation 
\begin{equation}
\frac{\partial h}{\partial t}+\nabla \vec{J}=0\;,
\label{eq:continuity1}
\end{equation}
to account for the surface evolution. Here
$h$ is the surface height and $\vec{J}$ is the current density of
diffusing adatoms. This equation is of course correct and reduces the
problem
to calculating $\vec{J}$.
It is widely assumed \cite{HagerSpohn,OzdemirZangwill} that $\vec{J}$
is
proportional to the gradient of the surface chemical potential
$\mu$, as expected in diffusive systems. In the case of
attachment-detachment limited systems
it was suggested by Nozieres \cite{Nozieres} that the chemical potential
gradient should be 
divided by the profile slope. Our model is consistent with the first
picture in the DL case and with
the second in the ADL case.

To show this we take the gradient of Eq.\ 
(\ref{eq:continuity1}). This leads to 
\begin{equation}
\frac{\partial \vec{D}}{\partial t}\propto \nabla \nabla \cdot
\vec{J}\;,
\label{eq:continuity2}
\end{equation} 
where $-\vec{D}$ is the gradient of the profile. We now 
return to Eq.\ (\ref{eq:differentialeq}) and transform
it back to an equation  
for the step density $D(\rho,\tau)$. Remembering that the term
$-\frac{xF'}{4}$ in  
Eq.\ (\ref{eq:differentialeq}) arises from the 
time derivative of $D$, we find that in the general case we can write 
\begin{eqnarray}
\frac{\partial D}{\partial \tau}
&+& \frac{\partial}{\partial \rho} \left( 
     \frac{1}{\rho} \frac{\partial}{\partial \rho} \left(
          \frac{\rho}{1-q+2qD} \cdot \frac{\partial \mu}{\partial \rho}
          \right)
\right)=0\;\;\; \mbox{with,}  
          \nonumber \\
\mu&=&\frac{1}{\rho}+g\left(\frac{D^2}{\rho}+3D\frac{\partial
D}{\partial \rho} \right)\;. 
\label{eq:DL_compact_eq}
\end{eqnarray}
Eq.\ (\ref{eq:DL_compact_eq}) is nothing but the radial component of
Eq.\
(\ref{eq:continuity2}) written in terms of the dimensionless variables
$\rho$
and $\tau$ with
\begin{equation}
J=\frac{1}{1-q+2qD}\cdot \frac{\partial \mu}{\partial \rho}\;.
\end{equation}

Therefore in the DL case ($q=0$) the adatom current is indeed 
proportional to the chemical potential gradient. In the ADL case ($q=1$)  
the chemical potential gradient is divided by the local step density as
suggested by Nozieres.
In addition we note that in
the limit $\rho  
\rightarrow \infty$ (where the steps are nearly straight), our chemical
potential (Eq.\ 
(\ref{eq:DL_compact_eq})) becomes identical to the chemical potential of
Ref.\ \onlinecite{HagerSpohn}.

\section{Properties of the scaling function}

In this section we use Eq.\ (\ref{eq:differentialeq}) to study
properties of the scaling function $F$. We also discuss the
boundary conditions necessary to solve Eq.\ (\ref{eq:differentialeq}).  
We begin by noting that according to the numerical simulations of the
step model, there is
a growing plateau or facet at the top of the cone. 
Our numerical solutions for the scaling function indicate that all
physical solutions indeed have such a special point, which can be
identified as the facet edge.
Let us examine the
behavior of the step
density on the facet and at its edge. At any given time there is only
one single step on the facet (the first step during its collapse
towards the origin). We have seen that the size of the facet grows
indefinitely,
and therefore the step density on the facet vanishes in the long time
scaling limit. Moreover, in Appendix C we show that in the long time
limit, the step density is a continuous function even at the facet edge;
i.e., it goes to zero continuously as the facet edge is approached from
above.

Denoting the scaled position of the facet edge by $x_0$, the above
observations can be expressed in the form 
\begin{equation}
  \label{eq:bc2}
  F(x)=0 \;,\; \forall x \leq x_0 \;.
\end{equation}
Note that $F(x)=0$ is not a solution of Eq.\
(\ref{eq:differentialeq}). This does not contradict the continuum
model, since the model was derived only for the case of a
finite step density (see Eq. (\ref{eq:xsdifference})). Thus the scaling
function has to satisfy Eq.\ (\ref{eq:differentialeq}) only at $x>x_0$,
and $x=x_0$ is a {\em singular} point.

Let us now study
the nature of this singularity at the facet edge.
Returning to our simulation data we
note that near the facet edge, the scaling function is extremely steep
(see Fig.\ \ref{scaling}). Indeed, the expansion of
Eq.\ (\ref{eq:differentialeq}) in small $F$ suggests that in the
vicinity of $x_0$, $F(x)$ can be written as a power series in
$\sqrt{x-x_0}$. Thus although $F$ is continuous at 
$x_0$, all its derivatives with respect to $x$ diverge at the
singular point.

We now turn to discuss the
boundary conditions necessary to solve Eq.\ (\ref{eq:differentialeq}).
First, we examine the behavior of $F$ at infinity.
Far enough from the origin the steps do not move. Hence in the limit
$\rho\rightarrow\infty$ the step density remains at its initial
value. This implies that
\begin{equation}
  \label{eq:bc1}
  \lim_{x\rightarrow\infty}F(x)=1\;.
\end{equation}
It turns out that this condition corresponds to two boundary conditions,
since it is only possible to satisfy Eq.\ (\ref{eq:bc1}) if $F'$
vanishes at infinity. In fact, by considering the asymptotic expansion
of $F$ it can be shown that for large $x$
\begin{equation}
\label{eq:asymptotic_expansion}
F=1+{\em a}_4x^{-4}+O\left(x^{-8}\right)\;,
\end{equation}
where ${\em a}_4=3\left(1+g\right)$ in the DL case and ${\em
a}_4=\frac{3}{2}\left(1+g\right)$
in the ADL case.

In order to solve Eq.\ (\ref{eq:differentialeq})
we need two additional boundary conditions at $x=x_0$.
The first of these is $F(x_0)=0$ (see Eq.\ (\ref{eq:bc2})).
Another boundary condition at $x=x_0$ can be derived by enforcing volume
conservation. From Eq.\ (\ref{eq:volume}), the volume change during the  
time $\tau$ is given by
\begin{equation}
  \label{eq:volumechange1}
  2\pi\int_0^\infty \rho \left[ h(\rho,\tau)-h(\rho,0) \right] d\rho\;.
\end{equation}
Integrating this equation by parts and requiring volume conservation, we
get
\begin{equation}
  \label{eq:volumechange2}
  \int_0^\infty \rho^2  \left[ D(\rho,\tau)-D(\rho,0) \right] d\rho=0\;.
\end{equation}
In terms of scaled variables the last equation can be written as
\begin{equation}
  \label{eq:volumechange3}
  \int_0^\infty x^2 \left( F-1 \right)dx=\int_{x_0}^\infty x^2 \left(
  F-1 \right)dx-\frac{x_0^3}{3}=0\;,
\end{equation}
where we have used the fact that $F$ vanishes below $x_0$.
The evaluation of the integral in Eq.\ (\ref{eq:volumechange3}) can be
done by multiplying Eq.\
(\ref{eq:differentialeq}) by $x^2$ and integrating it with
respect to $x$. The resulting equation is 
\begin{equation}
  \label{eq:localconservation1}
  \int_{x_0}^{\infty}x^2\left(\frac{F'{\cal A}}{2}+F^2{\cal B}
  \right)dx=
  \int_{x_0}^{\infty}\frac{x^3F'}{4}dx\;.
\end{equation}
The integral ${\cal M}_2=\int x^2\left(\frac{F'{\cal A}}{2}+F^2{\cal B}
\right)dx$ on the l.h.s.\ is carried out in appendix C, and is expressed
in terms of $F$ and its derivatives. The result of
this
integration combined with integration by parts of the r.h.s.\ of Eq.\
(\ref{eq:localconservation1}) leads to  
\begin{equation}
  \label{eq:localconservation2}
  \int_{x_0}^{\infty}x^2(F-1)dx=\left. \frac{1}{3}\left[x^3(F-1)-4{\cal
  M}_2\right] \right|_{x_0}^{\infty}\;.
\end{equation}
Inserting this relation in 
Eq.\ (\ref{eq:volumechange3}) 
we obtain the boundary condition
\begin{equation}
  \label{eq:bc3}
\left. {\cal M}_2 \right|_{x_0}=0\;.
\end{equation}
We have used the facts that the surface term at infinity in Eq.\
(\ref{eq:localconservation2})
vanishes and $F(x_0)=0$.
 
At this point we have four boundary conditions, two at infinity and
two at $x_0$. We may now obtain a unique solution of Eq.\
(\ref{eq:differentialeq}) if we know the value of $x_0$. What
determines $x_0$? To answer this question,
consider the height of the cone at the origin, $h_0$, at time
$\tau=\tau_{n-1}$ and at time $\tau=\tau_{n}$. $\tau_{n-1}$
and $\tau_n$ are the disappearance times of two successive
steps, and therefore 
$h_0\left(\tau_{n-1}\right)-h_0\left(\tau_n\right)=1$. We can also
calculate this difference from Eq.\ (\ref{eq:hzero}). Combining these
two results we arrive at the relation
\begin{equation}
\left(\tau_n^{1/\gamma}-\tau_{n-1}^{1/\gamma}\right)\int_0^\infty\left( 
F-1\right)dx=-1\;.
\end{equation}
Recalling that $\tau_n=\left(\theta_0 n\right)^\gamma$ and using the
fact that 
$F$ vanishes below $x_0$, we can rewrite the last equation as
\begin{equation}
\label{eq:height_conservation}
\int_{x_0}^\infty\left( F-1\right)dx-x_0=-\theta_0^{-1}\;.
\end{equation}
The integral in Eq.\ (\ref{eq:height_conservation}) can
be evaluated by integrating Eq.\ (\ref{eq:differentialeq}) with respect
to $x$. The result is:
\begin{equation}
\label{eq:height_integral_eval}
 \int_{x_0}^{\infty}\left( F-1 \right)dx=\left. 4{\cal M}_0
 \right|_{x_0}+x_0\;,
\end{equation}
where the integral ${\cal M}_0=\int \left(\frac{F'{\cal A}}{2}+F^2{\cal
B}
\right)dx$ is carried out in Appendix C, and is expressed in terms of
$F$ and its derivatives. Combining Eq.\
(\ref{eq:height_conservation}) and 
Eq.\ (\ref{eq:height_integral_eval}) we obtain the relation
\begin{equation}
\label{eq:theta_zero_relation}
\left. 4{\cal M}_0 \right|_{x_0}=-\theta_0^{-1}\;.
\end{equation}
The l.h.s.\ of this equation depends on $x_0$, thus relating
$x_0$  to $\theta_0$, the scaled
collapse time period of the steps. 
What is then the value of $\theta_0$ ? It is determined by the motion
and collapse of the first step. The behavior of the first step is
different from that of all the other steps, since it does not have
neighboring steps with smaller radii. Thus, our continuum model, which
treats all the steps on equal footing, does not contain any information
on the value of $\theta_0$. We therefore expect {\em a family} of valid
scaling functions consistent with the continuum model with different
values of $x_0$ or $\theta_0$. The unique value of $x_0$ observed in
simulations is determined by the discrete nature of the steps, and at
this stage we are not able to calculate it.

\section{Numerical evaluation of the scaling function}

We now find the scaling function numerically. Consider first the DL
case. We choose a value of $x_0$ and solve equation
(\ref{eq:differentialeq}) starting
at $x=x_0$. As explained in the previous section, $F$ can be expanded in
powers of $\sqrt{x-x_0}$ in the vicinity of $x_0$:
\begin{equation}
  \label{eq:root_expansion}
  F(x)=\sum_{n=1}^\infty k_n \left( \sqrt{x-x_0}\; \right) ^n \;,
\end{equation}
where we have already used the boundary condition (\ref{eq:bc2}). We
therefore have three additional free parameters in this expansion, which
should be determined by the boundary conditions (\ref{eq:bc1}) and
(\ref{eq:bc3}). We choose these parameters as the first three odd
coefficients in the expansion (\ref{eq:root_expansion}): $k_1$, $k_3$
and $k_5$ (the first two even coefficients vanish). Using
the boundary condition (\ref{eq:bc3}), we can express $k_5$ as a
function of $k_1$ and $k_3$ through the relation
\[
 k_5=\frac{k_1}{6x_0^2} -\frac{k_3^2}{2k_1} +\frac{k_3}{9x_0}
 -\frac{1}{6gx_0^3k_1}\;,
\]
thus reducing the number of free parameters to two. 

Finally we use the boundary condition (\ref{eq:bc1}) to find the
coefficients $k_1$ and $k_3$. Since (\ref{eq:bc1}) addresses the value
of $F$ at infinity, it cannot be easily applied to the expansion of $F$
in the vicinity of $x_0$. We therefore start from an initial guess for
the values of $k_1$ and $k_3$, solve Eq.\ (\ref{eq:differentialeq})
numerically for this choice of parameters and then tune $k_1$ and $k_3$
until the boundary condition (\ref{eq:bc1}) is satisfied. For a given
choice of $k_1$ and $k_3$, we found the solution of Eq.\
(\ref{eq:differentialeq}) in the following way. Numerical integration
starting at $x_0$ is impossible because the derivatives of $F$ diverge
there. Therefore, we first used the expansion (\ref{eq:root_expansion}),
truncated at a high enough order, to evaluate the function $F$ and its
first three derivatives at $x=x_0+\delta x$, for some choice of $\delta
x$. Then, using these values we integrated Eq.\
(\ref{eq:differentialeq}) numerically from $x_0+\delta x$. We made sure
that the solution is not sensitive to the choice of $\delta x$.  

\begin{figure}[h]
\centerline{
\epsfxsize=80mm
\epsffile{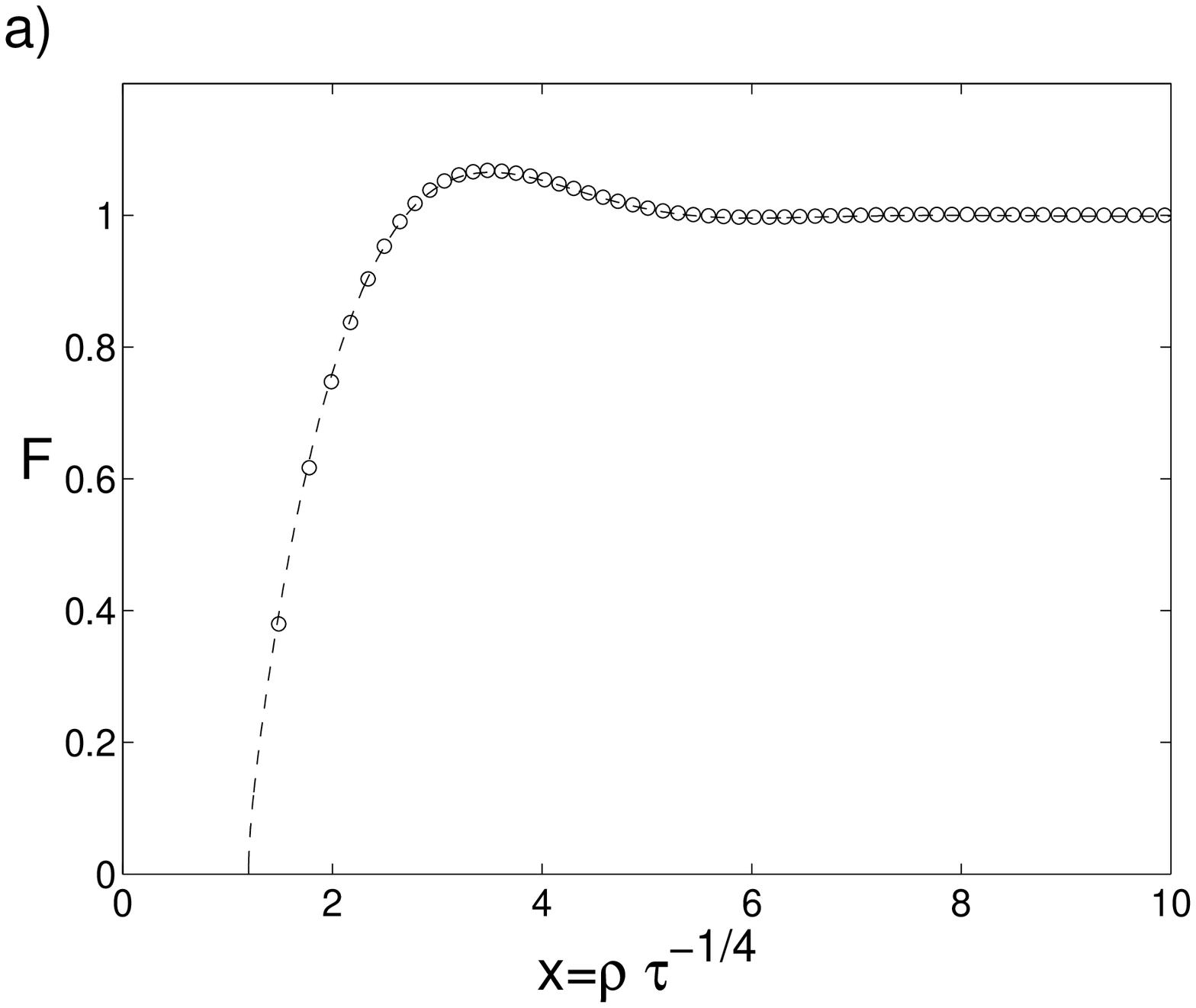}  
\hspace{0.1in}
\epsfxsize=80mm
\epsffile{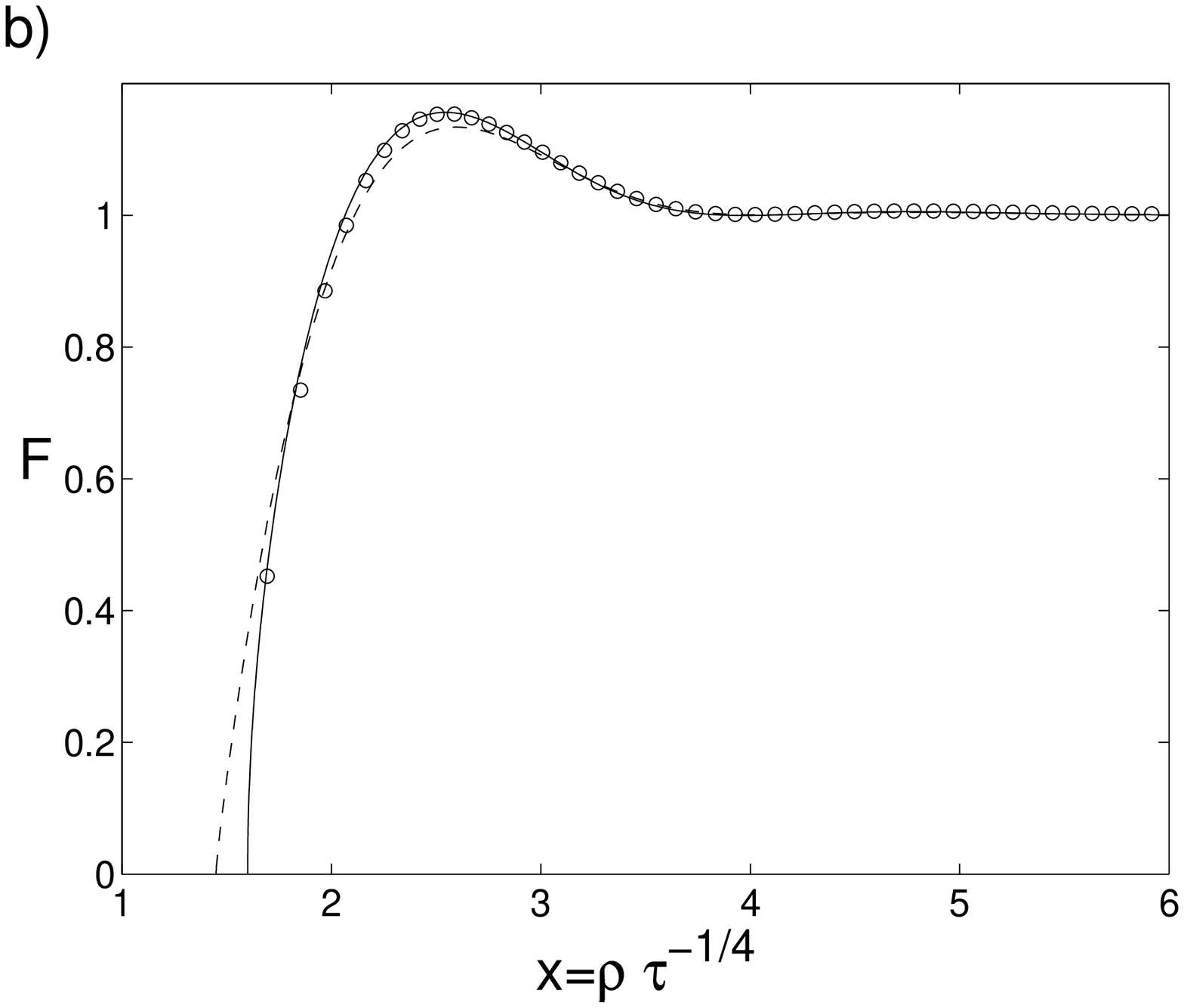}}
\vspace{0.1in}
\centerline{
\epsfxsize=80mm
\epsffile{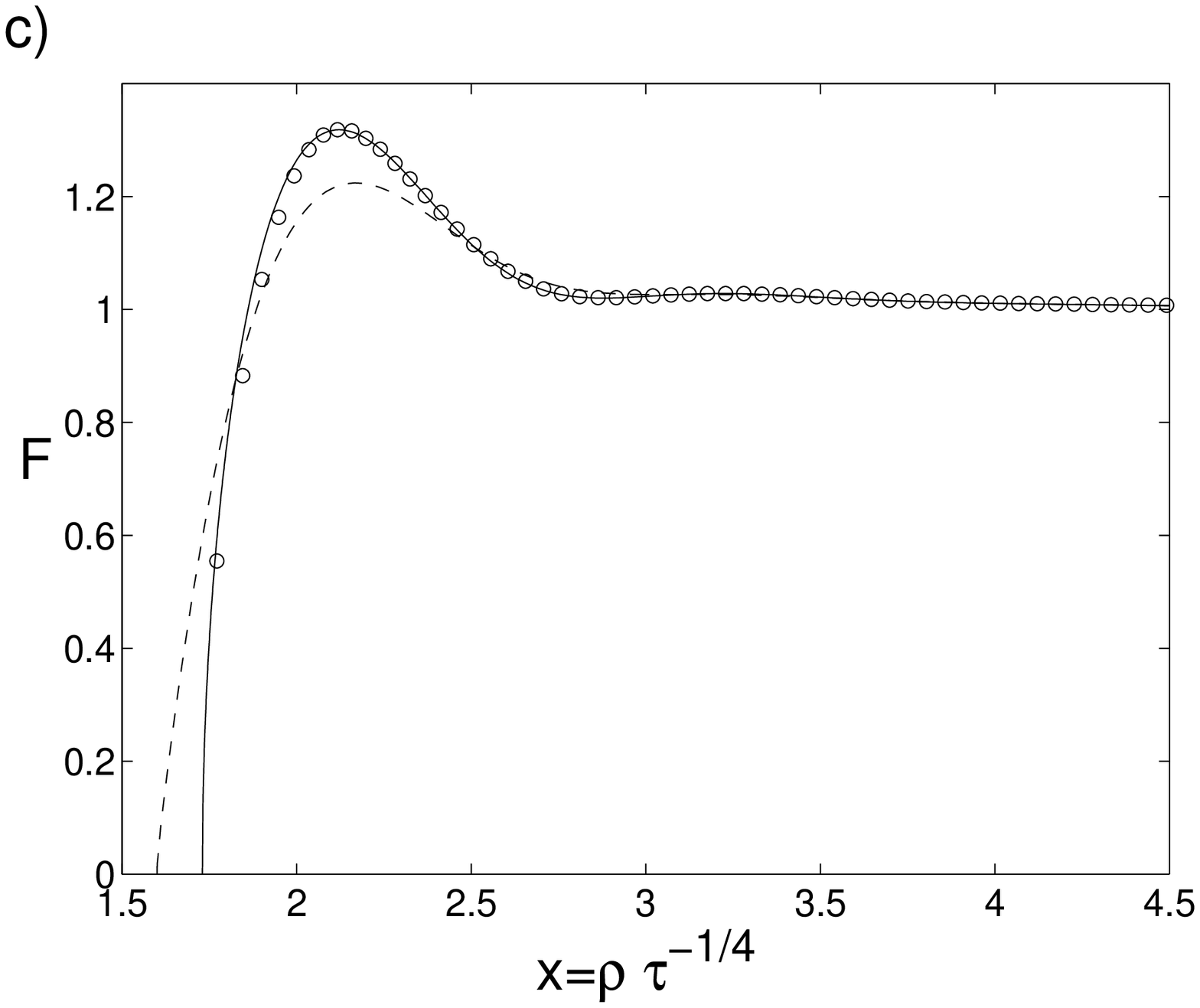}} 
\vspace{0.1in}
\caption{Numerical solutions for the DL scaling function compared
with simulation  
data. Results for three different values of the step-step interaction
parameter are
shown:
 (a) $g=0.1$, (b) $g=0.01$ and (c) $g=0.001$. When the step-step
interaction is strong, 
  the $x_0^*$ solution (dashed line) agrees very well with the
  simulation
data (circles). As the value of  
  $g$ is reduced, the best fit solution of 
Eq.\ (\protect \ref{eq:differentialeq})  
  (solid) deviates from the $x_0^*$ solution to higher values of
$x_0$.}   
\label{DL_sim_vs_cont} 
\end{figure} 

The above procedure was employed to generate the scaling function for a
range of values of the scaled facet edge position, $x_0$, and the
following picture emerged. There is a special minimal value of $x_0$,
which we denote by $x_0^*$. For $x_0<x_0^*$ there is no solution of Eq.\
(\ref{eq:differentialeq}) which satisfies the boundary conditions. For
each value of $x_0 \geq x_0^*$ there is a unique solution which
satisfies
the boundary conditions. Thus, as we anticipated, there is a one
dimensional family of scaling functions parameterized by the value of
$x_0 \geq x_0^*$. 

Since $x_0$ is related to $\theta_0$ through Eq.\
(\ref{eq:theta_zero_relation}), this family of solutions can
alternatively be labeled by the value of $\theta_0$. We used Eq.\
(\ref{eq:theta_zero_relation}) to calculate $\theta_0(x_0)$ and found
that it is a monotonically decreasing function. In particular,
$\theta_0(x_0)$ is maximal
at $x_0^*$. This is reasonable, since it means that a smaller facet
corresponds to a longer period and thus to a slower facet edge.

Despite the existence of many possible scaling functions, our
simulations suggest that
the system reaches a unique scaling solution independent of initial
conditions. What is the selected solution? In Figs.\
\ref{DL_sim_vs_cont}
we compare
the calculated scaling functions with simulation data in the DL case.
We do this for 
three different values of $g$, the interaction strength parameter.
When $g$ is large there is an impressive agreement between the
simulations data and the $x_0^*$ solution (Fig \ref{DL_sim_vs_cont}
(a)).
When $g$ is reduced (i.e., for weaker interactions between steps)
the observed scaling function deviates from the $x_0^*$ solution (Fig.\
\ref{DL_sim_vs_cont} (b) and (c)). 
However, in these cases there is another solution 
with a larger value of $x_0$ that best fits the simulation data. The
agreement between 
this best fit solution and the simulation data is again excellent.

The above observations suggest that $x_0^*$ is the selected
solution in the large $g$ limit. 
We propose the following argument to support this scenario.
Since the parameter $g$ is a measure of the strength of the step-step
interaction $G$ relative to the step line  
tension $\Gamma$, the large $g$ limit is equivalent to the small
$\Gamma$ limit.  
The collapse of the first step is driven by the step line tension. 
In the large $g$ limit the 
collapse driving force is minimal and the collapse time period is
maximal. As we mentioned above,  
a long collapse period is equivalent to a large value of $\theta_0$,
which
corresponds to a small value of $x_0$. Thus  
the large $g$ limit corresponds to the minimal value of $x_0$, i.e.\
$x=x_0^*$. 

\begin{figure}[h]
\centerline{
\epsfxsize=80mm
\epsffile{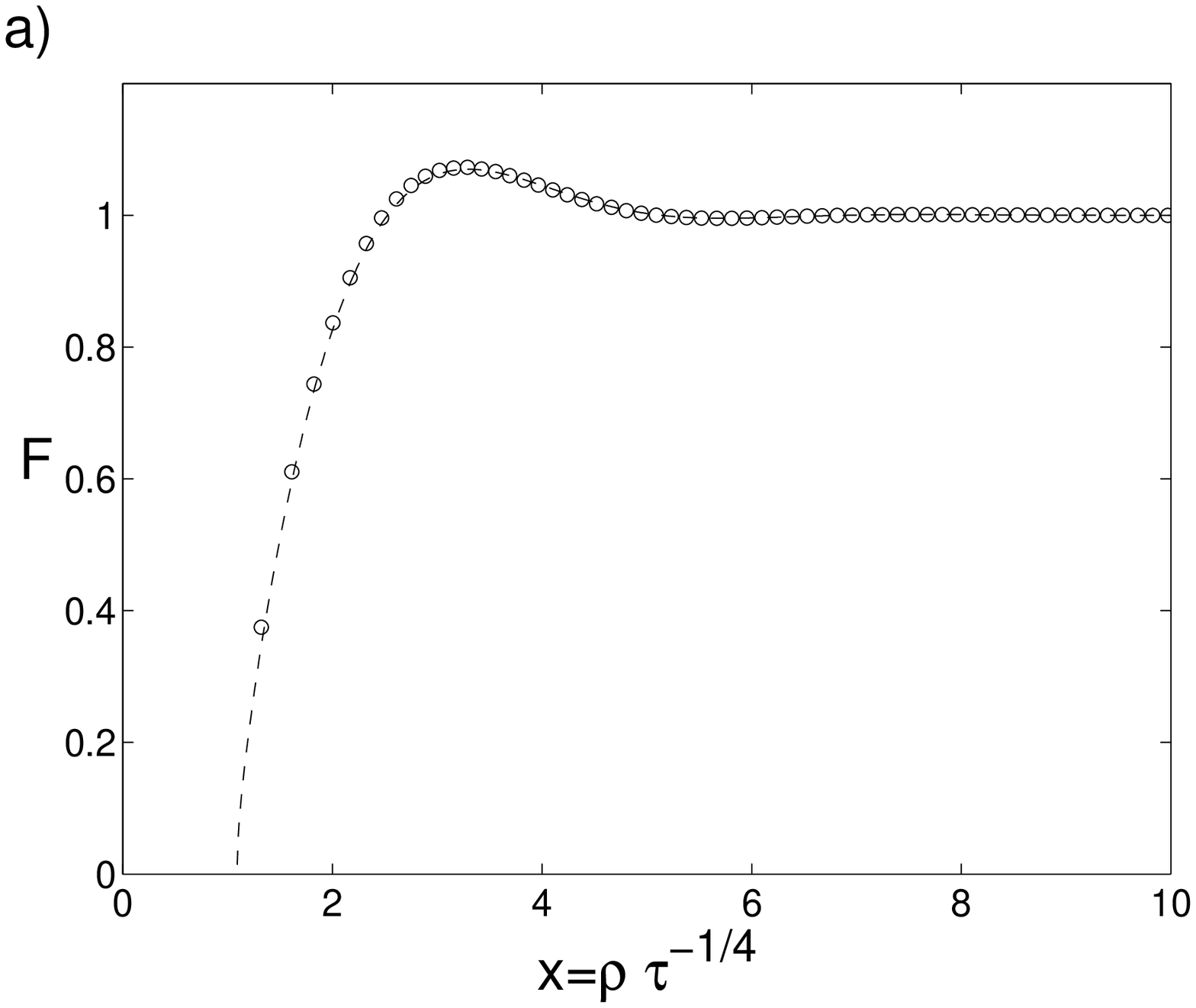}  
\hspace{0.1in}
\epsfxsize=80mm
\epsffile{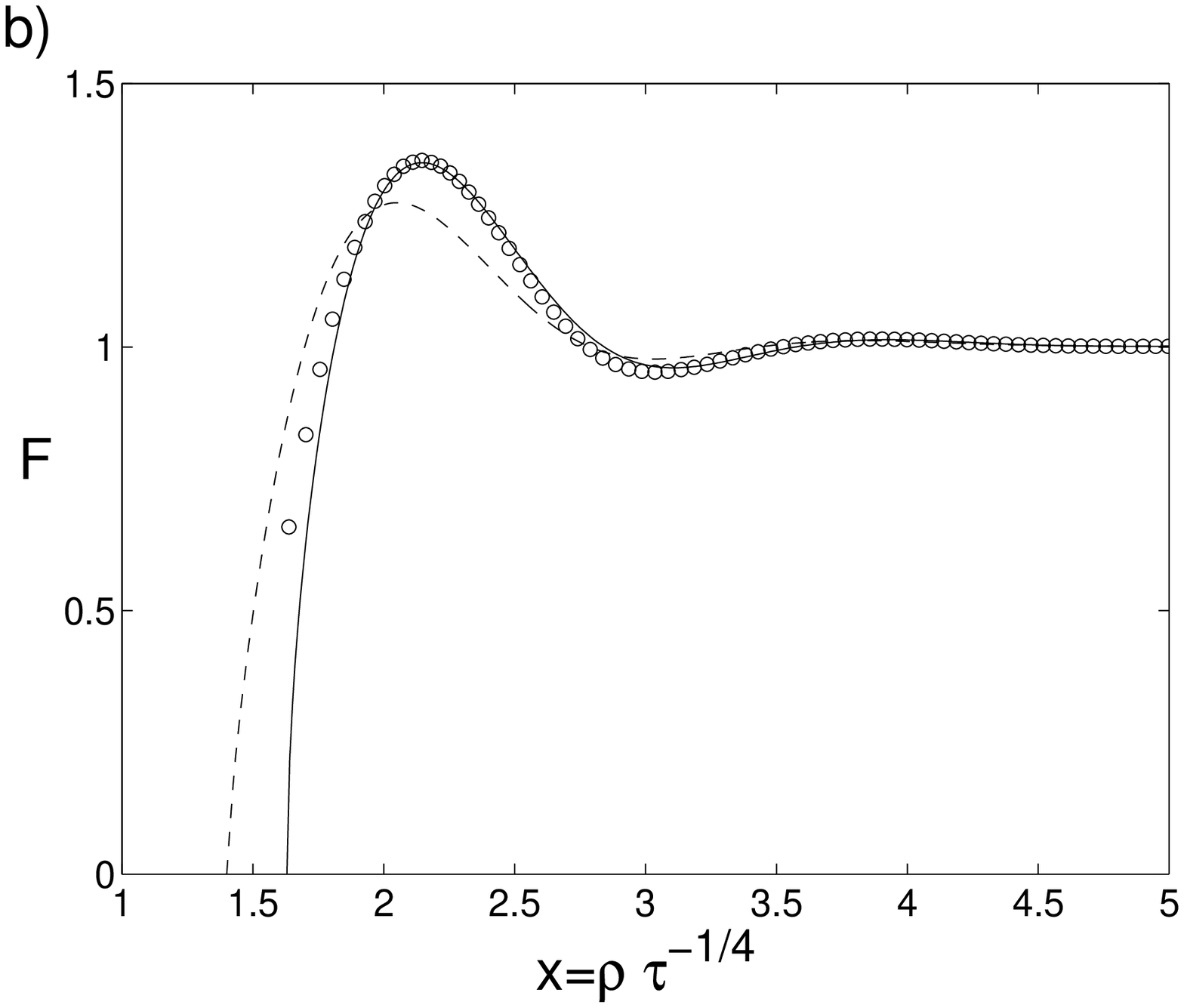}}
\vspace{0.1in}
\caption{Numerical solutions for the ADL scaling function compared
with simulation  
data. Results for two different values of the step-step interaction
parameter are
shown:
 (a) $g=0.2$ and (b) $g=0.01$. When the step-step
interaction is strong the simulation data (circles) agrees with the
$x_0^*=1.09 \pm 0.07$ solution (dashed line). 
As $g$ is reduced, we notice that the best fit solution of 
Eq.\ (\protect \ref{eq:differentialeq}) (solid line) deviates from
the $x_0^*=1.4 \pm 0.07$ solution to
higher values of 
$x_0$.}   
\label{ADL_sim_vs_cont} 
\end{figure} 

Now consider the ADL case. Although in this limit we can also expand
$F$ in $\sqrt{x-x_0}$, the numerical procedure described above is not an
effective method of solution in this case. It turns out that the
resulting scaling functions are sensitive to the choice of $\delta x$. 
We therefore had to use a
different method to solve Eq.\ (\ref{eq:differentialeq}) in the ADL
case. Let us denote by $x_{peak}$ the minimal value of $x$, which
corresponds to a local maximum of $F$. Such a point must exist for any
continuous function
satisfying the integral condition  
(\ref{eq:volumechange3}) and  
the boundary 
conditions (\ref{eq:bc2}) and (\ref{eq:bc1}). 
By tuning the values of
$F(x_{peak})$, $F''(x_{peak})$ and $F'''(x_{peak})$ we found a one
dimensional family of scaling functions, which satisfy the boundary
conditions. This family can be parameterized by the value of $x_{peak}$
or alternatively by the value of 
the resulting $x_0$. As in the DL case, we find that there is a minimal
value of $x_0$, denoted by $x_0^*$, below which there are no solutions
satisfying the boundary conditions.

Figs. \ref{ADL_sim_vs_cont} show the simulation data compared
with calculated scaling functions for two values of the interaction
strength $g$. We again find that for large $g$ the simulation
data agrees with 
the $x_0^*$ solution (dashed line). When $g$ is reduced, the agreement
deteriorates and  
there is a different solution with $x_0>x_0^*$ (solid line) which best
fits the simulation data.

\section{Effects of step permeability}

The step flow model we introduced in section II assumes that steps are 
impermeable; i.e., adatoms cannot hop between neighboring terraces
without being incorporated in the step separating them. In general,
however, steps may be permeable. 
For example, in a recent paper\cite{TanakaBartelt}, Tanaka et al.\
interpret their
experimental results as evidence that steps on Si(001) are permeable. We
therefore ask the following question: What is the effect of step
permeability on decay of nanostructures in general, and on the decay of
an infinite cone in particular? In this section we show that the scaling
exponents of an infinite cone of permeable steps are identical to the
exponents associated with impermeable steps. Moreover, the only effect
of step permeability on the differential equation for the scaling
function is a renormalization of one of its parameters.

Following Ref.\ \onlinecite{TanakaBartelt} we generalize our step flow
model to include
step permeability. We assume that flux of adatoms between two
neighboring terraces due to direct hops is determined by first order
kinetics. Introducing the permeability coefficient $p$, we rewrite
Eq.\ (\ref{eq:boundry}) as
\begin{eqnarray}
D_s \left. \frac{\partial  C_i }{\partial r}
\right|_{r_i}&=&k\left(\left.C_i
  \right|_{r_i}-C^{eq}_i\right)+p\left(\left.C_i\right|_{r_i}-\left.C_{i
  -1}\right|_{r_i}\right)\; \nonumber \\
-D_s \left. \frac{\partial C_{i-1} }{\partial r}
\right|_{r_i}&=&k\left(\left.C_{i-1}
  \right|_{r_i}-C^{eq}_i\right)-p\left(\left.C_i\right|_{r_i}-\left.C_{i
  -1}\right|_{r_i}\right)\;.
\label{eq:perma_boundry}
\end{eqnarray}
Step permeability does not affect the diffusion equation itself, and
therefore the general form of the diffusion field, given by Eq.\
(\ref{eq:diff_solution}), remains unaltered. However the coefficients
$A_i$ and $B_i$ are affected, and the equations for coefficients of
different terraces become coupled. As a result, the step equations
of motion cannot be written in closed form. The scaling
analysis therefore becomes more cumbersome.

As in the case of impermeable steps we assume that in the long time
limit the scaling ansatz
Eq.\ (\ref{eq:scaling}) holds, with the same definitions of the scaled
variables, $x$ and $\theta$ in terms of $\rho$ and $\tau$. We slightly
change the definition of the dimensionless time $\tau$ to $\tau=t/t_0$,
where $t_0$ is a time scale which
we can choose to our convenience. 
The derivation of the exponents $\alpha$ and $\beta$ in section IV did
not
involve the dynamics of the system. Thus the same derivation still holds
and the values of $\alpha$ and $\beta$ are not affected by step
permeability. We proceed to evaluate the exponent $\gamma$ and the
scaling function $F$ by studying Eq.\
(\ref{eq:perma_boundry}). 

For convenience, we parametrize the solutions of the diffusion equation
by concentration differences instead of the parameters
$A_i$ and $B_i$. These concentration differences, $U_i$ and $V_i$, are
defined as follows:  
\begin{eqnarray}
U_i&=&C_i(r_i)-\bar{C}^{eq}=A_i\ln{r_i}+B_i-\bar{C}^{eq}\; \nonumber \\
V_i&=&C_i(r_i)-C_i(r_{i+1})=A_i\ln{\frac{r_i}{r_{i+1}}} \; .
\label{eq:conc_diffs}
\end{eqnarray}
In the continuum limit, $U_i$ and $V_i$ are continuous
functions of $r$ and $t$ or alternatively of $x$ and $\theta$. We 
associate the values of $U_i$ and $V_i$ with the middle of the $i$th
terrace, namely
\begin{eqnarray}
U\left(\frac{x_i+x_{i+1}}{2},\theta \right)&=&U_i\;, \nonumber \\
V \left(\frac{x_i+x_{i+1}}{2},\theta \right)&=&V_i \;.
\label{eq:two_fields}
\end{eqnarray}
The scaling scenario for the functions $U$ and $V$ is
\begin{eqnarray}
U\left(x,\theta\right)&=&\theta^\mu\left(u\left(x\right)+O\left(\theta^{
-1}\right)\right)\;,
\nonumber \\
V\left(x,\theta\right)&=&\theta^\nu\left(v\left(x\right)+O\left(\theta^{
-1}\right)\right)\;.
\label{eq:two_fields2}
\end{eqnarray}
In terms of $U_i$, $V_i$ and the dimensionless radii $\rho_i$, Eq.\
(\ref{eq:perma_boundry}) takes the
form:
\begin{eqnarray}
\frac{T D_s
  V_i}{\Omega \Gamma
  \rho_i\ln{\frac{\rho_i}{\rho_{i+1}}}}&=&k\left(U_i-\bar{C}^{eq}\xi_i
  \right)+p\left(U_i-\left(U_{i-1}-V_{i-1}\right)\right)\;, \nonumber \\
-\frac{T D_s
  V_{i-1}}{\Omega \Gamma
  \rho_i\ln{\frac{\rho_{i-1}}{\rho_i}}}&=&k\left(U_{i-1}-V_{i-1}-\bar{C}
  ^{eq}\xi_i\right)-p\left(U_i-\left(U_{i-1}-V_{i-1}\right)\right)\;.
\label{eq:perma_boundry2}
\end{eqnarray}
 
Evaluating Eq.\ (\ref{eq:perma_boundry2}) at $x=(x_i+x_{i+1})/2$ we
can now employ Eq.\ (\ref{eq:xisexpansion}) to expand $\xi_i$,
$\rho_i\ln{\frac{\rho_i}{\rho_{i+1}}}$ and
$\rho_i\ln{\frac{\rho_{i-1}}{\rho_i}}$ in powers of
$\theta^{-1}$:
\begin{eqnarray}
&\xi&_i=\left( {\frac{1}{x}} + g\,\left( {\frac{{F^2}}{x}} +
{3\,F\,F'} \right)  \right) \,
   \theta^{-1} + O\left(\theta^{-2}\right)\;, \nonumber \\
&\rho&_i\ln{\frac{\rho_i}{\rho_{i+1}}}=-{\frac{1}{F}} +
O\left(\theta^{-1}\right)\;,
\label{eq:triv_expansion}\\
&\rho&_i\ln{\frac{\rho_{i-1}}{\rho_i}}=-{\frac{1}{F}} +
O\left(\theta^{-1}\right)\;. \nonumber
\end{eqnarray}
The $x_i$'s dependence of $U$ and
$V$ is unknown so we expand:
\begin{eqnarray}
u\left(\frac{x_{i-1}+x_i}{2}\right)&=&\sum_{n=0}^{\infty}\frac{1}{n!}
\frac{\partial^n
  u(x)}{\partial
  x^n}\left(\frac{x_{i-1}+x_{i}}{2}-x\right)^n\; \nonumber \\
v\left(\frac{x_{i-1}+x_i}{2}\right)&=&\sum_{n=0}^{\infty}\frac{1}{n!}
\frac{\partial^n
  v(x)}{\partial
  x^n}\left(\frac{x_{i-1}+x_{i}}{2}-x\right)^n\;.
\label{eq:fields_expansion}
\end{eqnarray}
Since the difference between successive $x$'s is of order
$\theta^{-1}$, Eq.\ (\ref{eq:fields_expansion}) is also an expansion
in this small parameter.

Using Eq.\ (\ref{eq:fields_expansion}) together with Eqs.\
(\ref{eq:two_fields}) and (\ref{eq:two_fields2}) we isolate $v(x)$ in
Eqs.\ (\ref{eq:perma_boundry2}), keeping only the lowest orders in
$\theta^{-1}$. 
\begin{eqnarray}
\theta^{1+\nu}v(x)&=&-\frac{\Omega \Gamma k
  \left(\theta^{1+\mu}u(x)-\bar{C}^{eq}\left(\frac{1}{x}+g\left(\frac{F^
  2}{x}+3FF'\right)\right)\right)}{D_s T F+p\Omega\Gamma}\; \nonumber
  \\
\theta^{1+\nu}v(x)&=&\frac{\Omega \Gamma k
  \left(\theta^{1+\mu}u(x)-\bar{C}^{eq}\left(\frac{1}{x}+g\left(\frac{F^
  2}{x}+3FF'\right)\right)\right)}{D_s T F+(p+k)\Omega\Gamma}\;.
\label{eq:isolate_v}
\end{eqnarray}
Since $v(x)$ cannot vanish identically, the two expressions above are
consistent only if $\nu <-1$, $\mu=-1$ and
\begin{equation}
u(x)=\bar{C}^{eq}\left[\frac{1}{x}+g\left(\frac{F^2}{x}+3FF'\right)
\right]\;.
\label{eq:u_of_x}
\end{equation}
Thus, both sides of Eqs.\ (\ref{eq:isolate_v}) decay in
time. We now subtract the second line of Eq.\ (\ref{eq:perma_boundry2})
from the first line and obtain the following equation:
\begin{equation}
\frac{D_s T}{\Omega \Gamma
  }\left(\frac{V_{i-1}}{\rho_i\ln{\frac{\rho_{i-1}}{\rho_i}}}+\frac{V_i}
  {\rho_i\ln{\frac{\rho_i}{\rho_{i+1}}}}\right)=
  \left(k+2p\right)\left(U_i-U_{i-1}+V_{i-1}\right)\;.
\label{eq:subtract_boundaries}
\end{equation}
Again we expand to lowest order in $\theta^{-1}$ and isolate
$v(x)$:
\begin{equation}
v(x)=-\theta^{-\left(2+\nu\right)}\frac{\Omega \Gamma \left(k+2p\right)}
{2D_s T F^2 + \Omega \Gamma \left(k+2p\right)F}u'(x)\;.
\label{eq:v_of_x}
\end{equation}
$v(x)$ does not depend on $\theta$. This implies that $\nu=-2$ and
$v(x)$ is proportional to $u'(x)$.

To finish the scaling analysis we return to Eq.\
(\ref{eq:partial3}) and find the leading orders in $\theta^{-1}$ of
$\dot{\rho}_{i+1}+\dot{\rho}_i$ and
$\dot{\rho}_{i+1}-\dot{\rho}_i$. Using Eqs.\ (\ref{eq:velocity}) and
(\ref{eq:conc_diffs}) we
find that
\begin{equation}
\dot{\rho}_i=\frac{D_s T^2}{\Omega
  \Gamma^2}t_0\left(\frac{V_i}{\rho_i\ln{\frac{\rho_i}{\rho_{i+1}}
  }}- 
\frac{V_{i-1}}{\rho_i\ln{\frac{\rho_{i-1}}{\rho_i}}}\right)\;.
\label{eq:Vi_velocity}
\end{equation}  
Putting together Eqs.\ (\ref{eq:two_fields2}) (\ref{eq:triv_expansion})
and (\ref{eq:Vi_velocity}) we
conclude that to lowest order in $\theta^{-1}$
\begin{eqnarray}
  \label{eq:expand_perma_velocities}
  \dot{\rho}_{i+1}+\dot{\rho}_i&=&\theta^{-3}({\cal
    A}_p+O(\theta^{-1}))\;, \nonumber \\
  \dot{\rho}_{i+1}-\dot{\rho}_i&=&\theta^{-4}({\cal
    B}_p+O(\theta^{-1}))\;.
\end{eqnarray}
${\cal A}_p$ and ${\cal B}_p$ are expressions involving $F$, $F'$,
$F''$,
$F'''$ and $F''''$. Note that the
orders of $\theta^{-1}$ in these expressions are identical to those in
the equivalent expressions in section IV (Eq.\
(\ref{eq:expandvelocities})). These orders are responsible for the
result $\gamma=4$ in the case of impermeable steps. Therefore, in the
permeable case we also have 
$\gamma=4$. 

After expanding $\dot{\rho}_{i+1}+\dot{\rho}_i$ and
$\dot{\rho}_{i+1}-\dot{\rho}_i$  we can use
Eq. (\ref{eq:partial3}) to obtain the differential equation for the
scaling function in the permeable case. However this
exercise is not necessary. We avoid it
by arguing that the resulting differential equation in the
permeable case is equivalent to the one in the impermeable case with 
renormalization of some of the parameters. To see this we note that the
treatment we presented here is valid also in the impermeable
case. Thus when $p=0$, the general differential
equation must be equivalent to the equation derived in
Section IV. In addition,
the attachment/detachment rate $k$ and the
permeability $p$ affect the step velocities only through the function
$v(x)$, which depends on $k$ and $p$ only trough the sum
$k+2p$. Therefore, $k$ and $p$ affect the differential equation itself
only trough the sum $k+2p$. We conclude that the scaling function in the
case of permeable steps is identical to the scaling function of
impermeable steps with $k+2p$ in the former replacing $k$ in the latter.

\section{Summary and discussion}

We have studied the relaxation of an infinite crystalline
cone below the roughening   
temperature, in terms of a step flow  
model. The model was solved numerically and two types of dynamical
evolutions were found. When the repulsive interactions are strong
enough, the decay of the cone proceeds through the collapse of the
innermost steps, one step at a time. Weak interactions lead to a step
bunching instability (except in pure diffusion limited kinetics), and
the decay process becomes much more complicated and involves collapse of
bunches of steps.

Focusing on stable cases, we found that in the long time
limit the decaying step system obeys
a scaling scenario. 
The step density (i.e. the slope of the height profile), defined as the
inverse step
separation, scales in time  
according to 
$D(\rho,\tau)=\tau^{\alpha/\gamma}F\left(\tau^{-\beta/\gamma}\rho,\tau^{
1/\gamma}\right)$.
$F$ is a function of the scaled position $x=\tau^{-\beta/\gamma}\rho$
and exhibits a periodic dependence on the scaled time
$\theta=\tau^{1/\gamma}$.
In particular, the position of the facet edge at the top of the cone
grows as $\tau^{\beta/\gamma}$.
The values of the scaling exponents which fit our simulations are
$\alpha=0$, $\beta=1$ and $\gamma=4$.   
 
Following this observation we used a scaling ansatz to transform the
discrete step flow model into a  
continuum description of surface evolution. The basic predictions
of this continuum  
model are the values of the scaling exponents (which agree with the
simulation results)  
and a differential equation 
for the scaling function $F$. This continuum model becomes exact in the
long time scaling limit, and it breaks down whenever the  
step density vanishes. This fact can be seen both from the
derivation of the  
continuum model and from the resulting differential equation, which
predicts a singular  
behavior at the zeros of the scaling function $F$.

We showed that each physical solution of the continuum equation has a
special point, $x_0$, at which the scaling function, $F$, vanishes. $F$
is singular at $x=x_0$, and we identify this special point as the scaled
position of a facet edge. Thus the  
edge of a macroscopic facet in the discrete system is a singular point
of the scaling function $F$.      

A detailed analysis of the discrete system revealed a sufficient number
of boundary conditions, which define a unique scaling function, $F$, for
a given value of the scaled facet edge position, $x_0$. However, we were
not able to find a unique value for $x_0$, and were left with a one 
dimensional family of solutions, parameterized by $x_0$.     
A numerical solution of the differential equation for $F$ confirmed the
existence of this family. We found that there is a minimal
value of $x_0$, which we denoted by  
$x_0^*$. For every $x_0 \geq x_0^*$ there is a unique scaling function,
while for $x_0<x_0^*$ there  
are no solutions, which satisfy the boundary conditions.

A comparison of the numerical solutions 
of $F$ with results of simulations of the discrete system leads to the
following picture. When the  
step-step repulsion is strong, there is a remarkable agreement between
the $x_0=x_0^*$ solution and  
the simulation data. When the magnitude of the step-step interaction is
reduced, the system reaches
scaling solutions   
with $x_0>x_0^*$. We therefore advance the hypothesis that in the strong
interaction limit the system approaches
the minimal $x_0^*$ solution.

Although our work provides a detailed account of the decay
process of an infinite cone, it leaves a few unresolved issues. First
and most important is the fact that we were not able to uniquely
determine the value of $x_0$. We did show that $x_0$ depends crucially
on the detailed collapse process of the innermost step, which is not
included in the continuum model. Thus, in order to evaluate $x_0$, one
has to deal directly with the discrete step system. Another open issue
related to the behavior of the innermost step is the
periodic behavior of the scaling function. This periodicity is observed
in the kinetics of the discrete system of steps, and is absent from the
solution of the continuum equation. Lastly, when the repulsive
interaction between steps is weaker than a certain threshold, the system
becomes unstable and step bunches are formed. The critical value of the
interaction below which this instability occurs and its dependence on
the kinetic parameter $q$ have not been studied so far. We intend to
address these open questions in future work. 

Finally, we remark that the scaling behavior of the cone profile,
predicted in this work, is robust. In particular, the
existence of a facet growing as $\tau^{1/4}$, as well the existence of a
scaling state does not depend on the detailed form of the repulsive
interactions between steps. A quantitative change of these interactions
may alter the scaling function, $F$, but the scaling exponents do not
change. Another manifestation of the robustness of the scaling solution
is the effect of step permeability. In principle, step permeability
could have changed the
scaling behavior of the system entirely. However, we showed
that its only effect on the scaling solution is to
modify the attachment-detachment rate coefficient, $k$, to $k+2p$.

We are greatful to N. Bartelt, H. C. Jeong, D. J. Liu and J. D. Weeks
for helpful discussions.
This research was supported by grant No. 95-00268 from the
United States-Israel Binational Science Foundation (BSF), Jerusalem,
Israel.
D. Kandel is the incumbent of the Ruth Epstein Recu
Career Development Chair.

\appendix
\section{}

Here we study the linear stability of circular, uniformly spaced
steps with unit
step separation in the absence of step-step interactions. This
configuration with $\rho_n=n$ is not a steady state. However, in
the
large radius limit the steps move very slowly, and the uniform state is
extremely close to a steady state. We therefore refer to it as a quasi
steady state. We regard the quasi steady state as unstable if the growth
of perturbations is faster than the motion of steps.
      
In the absence of step-step interactions Eqs.\
(\ref{eq:dimensionlessvelocity}) simplify to
\begin{eqnarray}
\dot{\rho}_n&\equiv&\frac{d\rho_n}{d\tau}=\frac{a_n-a_{n-1}}{\rho_n}\;,
\;\;
\mbox{with}  \label{eq:no_interaction_velocity} \\ 
a_n&=&\frac{\frac{1}{\rho_n}-\frac{1}{\rho_{n+1}}}
{\left(1-q\right)\ln \frac{\rho_n}{\rho_{n+1}} - 
q\left(\frac{1}{\rho_n} + \frac{1}{\rho_{n+1}}\right)} \nonumber \\
\end{eqnarray}
For large values of $n$, the velocity of the $n$th step in the quasi
steady state is  
given by: 
\begin{equation}
\dot{\rho}_n=\frac{n^{-3}}{1+q}+O \left( n^{-4} \right)\;.
\end{equation}
To check the linear stability of the above configuration
we perturb the step positions according to
\begin{equation}
\rho_n=n+\Delta e^{i \left(\phi n - \omega t\right)}\;.
\label{eq:perturbation}
\end{equation}

Equating the time derivative of this perturbation with the velocity of
the $n$th step results in an equation for $\omega$. 
We find that to the lowest orders in $\Delta$ and $n^{-1}$
\begin{equation}
\omega=\frac{4iq\left(1-\cos{\phi}\right)n^{-2}}{\left(1+q\right)^2}\;.
\end{equation}
Since the magnitude of $\omega$ is significantly larger than the step
velocities in 
the quasi steady state, positive values of $Im(\omega)$ lead to
an instability.
We see that away from the DL case ($q>0$) the system is
unstable. The most
unstable mode is $\phi=-\pi$ (step pairing).
The $\phi=0$ mode (uniform 
translation) is marginal as is the $q=0$ (DL) case.
 
\section{}

In this appendix we give some technical details of the algebraic
manipulations
performed in sections IV and V.

To calculate expressions ${\cal A}$ and ${\cal B}$ in Eq.\
(\ref{eq:expandvelocities}), we first express the scaled
positions of the steps as power series in  $\theta^{-1}$.
Inserting Eq.\ (\ref{eq:xisexpansion}) into Eq.\
(\ref{eq:taylorexpansion}) we find that to fifth order in $\theta^{-1}$
\begin{eqnarray}
x_{i-2}=&x& - {\frac{5\,\theta^{-1}}{2\,F}} -
{\frac{3\,F'\,{\theta^{-2}}}{{F^3}}} + 
  {\frac{5\,\left( -3\,{{F'}^2} + F\,F'' \right)
  \,{\theta^{-3}}}{2\,{F^5}}} + 
  {\frac{\left( -93\,{{F'}^3} + 61\,F\,F'\,F'' - 6\,{F^2}\,F^{(3)}
  \right) \,{\theta^{-4}}}{4\,{F^7}}} \nonumber \\
&-&{\frac{\left( 1935\,{{F'}^4} 
-1890\,F\,{{F'}^2}\,F'' + 260\,{F^2}\,F'\,F^{(3)} + 
        {F^2}\,\left( 180\,{{F''}^2} - 17\,F\,F^{(4)} \right)  \right)
  \,{\theta^{-5}}}{24\,{F^9}}} \nonumber \\ 
x_{i-1}=&x& - {\frac{3\,\theta^{-1}}{2\,F}} -
{\frac{F'\,{\theta^{-2}}}{{F^3}}} + 
  {\frac{\left( -3\,{{F'}^2} + F\,F'' \right)
  \,{\theta^{-3}}}{2\,{F^5}}} - 
  {\frac{\left( 33\,{{F'}^3} - 21\,F\,F'\,F'' + 2\,{F^2}\,F^{(3)}
  \right) \,{\theta^{-4}}}{12\,{F^7}}} \nonumber \\
&+& 
  {\frac{\left( -135\,{{F'}^4} + 126\,F\,{{F'}^2}\,F'' -
  12\,{F^2}\,{{F''}^2} - 
        16\,{F^2}\,F'\,F^{(3)} + {F^3}\,F^{(4)} \right)
  \,{\theta^{-5}}}{24\,{F^9}}} \nonumber \\
x_i\;\;\;=&x& - {\frac{\theta^{-1}}{2\,F}} \nonumber \\
x_{i+1}=&x& + {\frac{\theta^{-1}}{2\,F}} \nonumber \\
x_{i+2}=&x& + {\frac{3\,\theta^{-1}}{2\,F}} -
{\frac{F'\,{\theta^{-2}}}{{F^3}}} - 
  {\frac{\left( -3\,{{F'}^2} + F\,F'' \right)
  \,{\theta^{-3}}}{2\,{F^5}}} - 
  {\frac{\left( 33\,{{F'}^3} - 21\,F\,F'\,F'' + 2\,{F^2}\,F^{(3)}
  \right) \,{\theta^{-4}}}{12\,{F^7}}} \nonumber \\
&-& 
  {\frac{\left( -135\,{{F'}^4} + 126\,F\,{{F'}^2}\,F'' -
  12\,{F^2}\,{{F''}^2} - 
        16\,{F^2}\,F'\,F^{(3)} + {F^3}\,F^{(4)} \right)
  \,{\theta^{-5}}}{24\,{F^9}}} \nonumber \\
x_{i+3}=&x& + {\frac{5\,\theta^{-1}}{2\,F}} -
{\frac{3\,F'\,{\theta^{-2}}}{{F^3}}} - 
  {\frac{5\,\left( -3\,{{F'}^2} + F\,F'' \right)
  \,{\theta^{-3}}}{2\,{F^5}}} + 
  {\frac{\left( -93\,{{F'}^3} + 61\,F\,F'\,F'' - 6\,{F^2}\,F^{(3)}
  \right) \,{\theta^{-4}}}{4\,{F^7}}} \nonumber \\
&+& 
  {\frac{\left( 1935\,{{F'}^4} - 1890\,F\,{{F'}^2}\,F'' +
  260\,{F^2}\,F'\,F^{(3)} + 
        {F^2}\,\left( 180\,{{F''}^2} - 17\,F\,F^{(4)} \right)  \right)
        \,{\theta^{-5}}}{24\,{F^9}}}\;,
\end{eqnarray}
where $F$ and its derivatives are evaluated at
$x=\left(x_i+x_{i+1}\right)/2$.  

Using these expressions we expand the step velocities in $\theta^{-1}$
and obtain
Eq.\
(\ref{eq:expandvelocities}). In the general case the expressions for
${\cal A}$ and ${\cal B}$ are too cumbersome to be written
here. Instead we give these expressions in the two limiting cases. In
the DL case
\begin{eqnarray}
{\cal A}_{DL}&=&g\,\left( {\frac{2\,F}{{x^3}}} - {\frac{4\,F'}{{x^2}}} + 
     {\frac{10\,{{F'}^2}}{F\,x}} + {\frac{10\,F''}{x}} +
     {\frac{18\,F'\,F''}{F}} + 6\,F^{(3)} \right)+{\frac{2}{F\,{x^3}}}\;
     \nonumber \\
{\cal B}_{DL}&=&g\,\left( -{\frac{3}{{x^4}}} + {\frac{5\,F'}{F\,{x^3}}}
- {\frac{5\,{{F'}^3}}{{F^3}\,x}} - 
     {\frac{7\,F''}{F\,{x^2}}} + {\frac{10\,F'\,F''}{{F^2}\,x}} +
     {\frac{9\,{{F''}^2}}{{F^2}}} - 
     {\frac{{{F'}^2}\,\left( 5\,F + 9\,{x^2}\,F'' \right)
     }{{F^3}\,{x^2}}}\right. \nonumber \\
     &&\,\,\,\,\,\,\,\,\,\,\left. + 
     {\frac{5\,F^{(3)}}{F\,x}} + {\frac{9\,F'\,F^{(3)}}{{F^2}}} +
     {\frac{3\,F^{(4)}}{F}} \right)-{\frac{3}{{F^2}\,{x^4}}} -
     {\frac{F'}{{F^3}\,{x^3}}}\;.\label{ADL_AandB}
\end{eqnarray} 
Inserting Eq.\ (\ref{ADL_AandB}) into Eq.\ (\ref{eq:differentialeq}) we
find the
differential
equation which govern the scaling function in the DL case:
\begin{eqnarray}
&g&\, \left(12\,F'\,F^{(3)}+ 3\,F\,F^{(4)}+
  9\,{{F''}^2}+ {\frac{15\,F'\,F'' +
      5\,F\,F^{(3)}}{x}} \right. \nonumber \\
    &&\,\,\,\,\,- \left.{\frac{7\,\left( {{F'}^2} + F\,F'' \right)
    }{{x^2}}}+
    {\frac{6\,F\,F'}{{x^3}}}-{\frac{3\,{F^2}}{{x^4}}}\right)-{\frac{3}{{
    x^4}}} - {\frac{x\,F'}{4}}=0\;.
\end{eqnarray}

In the ADL case 
\begin{eqnarray}
{\cal A}_{ADL}&=&g\,\left(\frac{1}{x^3} - {\frac{F'}{F\,{x^2}}} +
{\frac{3\,{{F'}^2}}{{F^2}\,x}} - 
     {\frac{3\,{{F'}^3}}{{F^3}}} + {\frac{5\,F''}{F\,x}} +
     {\frac{6\,F'\,F''}{{F^2}}} + 
     {\frac{3\,F^{(3)}}{F}} \right)+{\frac{1}{{F^2}\,{x^3}}} +
     {\frac{F'}{{F^3}\,{x^2}}} \;, \nonumber \\
{\cal B}_{ADL}&=&g\,\left( -{\frac{3}{2\,F\,{x^4}}} +
{\frac{F'}{{F^2}\,{x^3}}} - {\frac{{{F'}^2}}{{F^3}\,{x^2}}} - 
     {\frac{3\,{{F'}^3}}{{F^4}\,x}} + {\frac{9\,{{F'}^4}}{2\,{F^5}}} - 
     {\frac{3\,F''}{{F^2}\,{x^2}}} + {\frac{F'\,F''}{2\,{F^3}\,x}} - 
     {\frac{21\,{{F'}^2}\,F''}{2\,{F^4}}} \right. \nonumber \\
   &&\,\,\,\,\,\,\left. + {\frac{3\,{{F''}^2}}{{F^3}}} + 
     {\frac{5\,F^{(3)}}{2\,{F^2}\,x}} +
     {\frac{3\,F'\,F^{(3)}}{2\,{F^3}}} + 
     {\frac{3\,F^{(4)}}{2\,{F^2}}} \right)-{\frac{3}{2\,{F^3}\,{x^4}}} -
     {\frac{2\,F'}{{F^4}\,{x^3}}} - 
  {\frac{3\,{{F'}^2}}{2\,{F^5}\,{x^2}}} +
  {\frac{F''}{2\,{F^4}\,{x^2}}}\;,
\end{eqnarray}   
and the differential equation for the ADL scaling function is:
\begin{eqnarray}
&g&\,\left( -{\frac{3\,F}{2\,{x^4}}} + {\frac{3\,F'}{2\,{x^3}}} - 
     {\frac{3\,{{F'}^2}}{2\,F\,{x^2}}} -
     {\frac{3\,{{F'}^3}}{2\,{F^2}\,x}} + 
     {\frac{3\,{{F'}^4}}{{F^3}}} - {\frac{3\,F''}{{x^2}}} +
     {\frac{3\,F'\,F''}{F\,x}} - 
     {\frac{15\,{{F'}^2}\,F''}{2\,{F^2}}} + {\frac{3\,{{F''}^2}}{F}} +
     {\frac{5\,F^{(3)}}{2\,x}} \right. \nonumber \\
&&\,\,\,\,\,\left. + 
     {\frac{3\,F'\,F^{(3)}}{F}} + {\frac{3\,F^{(4)}}{2}} \right)  
-{\frac{3}{2\,F\,{x^4}}} - {\frac{3\,F'}{2\,{F^2}\,{x^3}}} -
{\frac{x\,F'}{4}} - 
  {\frac{{{F'}^2}}{{F^3}\,{x^2}}} + {\frac{F''}{2\,{F^2}\,{x^2}}}=0\;.
\end{eqnarray}

In section V we used the moments
\begin{eqnarray}
 {\cal M}_0&=&\int \left(\frac{F'{\cal A}}{2}+F^2{\cal B}
\right)dx\;, \nonumber \\
 {\cal M}_2&=&\int x^2\left(\frac{F'{\cal A}}{2}+F^2{\cal B}
\right)dx\;. \nonumber 
\end{eqnarray}
to set boundary conditions for the scaling function at $x_0$.
 
In the DL case 
\begin{eqnarray}
{\cal M}_{0_{DL}}&=&g\left(
\frac{F^2}{x^3}-\frac{2F\,F'}{x^2}+\frac{5F'^2}{x}+\frac{5F\,F''}{x}+9F'
\,F''+3F\,F'''\right)+\frac{1}{x^3}\;,
\nonumber \\
{\cal M}_{2_{DL}}&=&g\left( \frac{3F^2}{x} - 6F\,F' - xF'^2 -xF\,F'' + 
     9x^2F'\,F'' + 3x^2F\,F''' \right) + \frac{3}{x}\;. \nonumber
\end{eqnarray}

In the ADL case
\begin{eqnarray}
{\cal M}_{0_{ADL}}&=&g\left( \frac{F}{2x^3} -\frac{F'}{2x^2} 
+\frac{3F'^2}{2xF} -\frac{3F'^3}{2F^2}+\frac{5F''}{2x}
+\frac{3F'\,F''}{F}+\frac{3F'''}{2} \right) \nonumber \\
  &\;&+\frac{1}{2x^3 F} + \frac{F'}{2x^2 F^2}\; \nonumber,
\nonumber \\
{\cal M}_{2_{ADL}}&=&g\left(\frac{3F}{2x} - \frac{5F'}{2} -
\frac{3xF'^2}{2F} - \frac{3 x^2 F'^3}{2 F^2} - \frac{xF''}{2} +
     \frac{3 x^2 F'\,F''}{F} + 
     \frac{3 x^2 F'''}{2} \right) \nonumber \\
   &\;&+\frac{3}{2 x F} + \frac{F'}{2F^2} \;. \nonumber 
\end{eqnarray}

\section{}

In this appendix we show that in the scaling state, the step density
near the 
facet edge must vanish when the facet size diverges. This implies that 
$F\left(x_0\right)=0$ where $x_0$ is the scaled position of the facet
edge.
We restrict ourselves to situations where steps collapse towards the
origin one at a time (consistently with simulation results). 

It is tempting to argue that since our system is expanding and slowing
down, in the 
long time limit it approaches the equilibrium state of straight steps
in contact with a facet. In this equilibrium state the step density
near the facet edge vanishes as a square root. This equilibrium state
is not consistent with our density function, which also predicts a
square
root approach to zero, but with a time dependent coefficient. Moreover
even in the long time limit steps in our system continue to
collapse. In each collapse period, just before the step disappears, its 
chemical potential diverges. At these times the system is not close to
equilibrium. 

Consider the
velocity of the first step. It depends on the positions of
the first three steps $\rho_1$, $\rho_2$ and $\rho_3$ through
\begin{equation}
  \label{eq:v1}
  \dot{\rho}_1=\frac{1}{\rho_1} \cdot
  \frac{\xi_1-\xi_2}{\left(1-q\right)\ln \frac{\rho_1}{\rho_2} - 
q\left(\frac{1}{\rho_1} + \frac{1}{\rho_2}\right)}\;.
\end{equation}
$\rho_2$ is always larger then $\rho_1$ so the denominator is always
negative. The direction of motion of the first step is thus given by
the sign of the numerator
\begin{equation}
  \label{eq:signv1}
  {\cal N}=\xi_1-\xi_2=\frac{1}{\rho_1}-\frac{1}{\rho_2}
  +2g\left(\frac{1}{\left(\rho_2-\rho_1\right)^3}-\frac{\rho_3}{\rho_2+
  \rho_3}\frac{1}{\left(\rho_3-\rho_2\right)^3}\right)\;,
\end{equation}
which must be positive to avoid bounding of the first and second
steps.

Fixing $\rho_2$ and $\rho_3$, the value of $\rho_1^*$ which minimizes
${\cal N}$ is found
by solving
\begin{equation}
\left. \frac{\partial {\cal N}}{\partial
\rho_1}\right|_{\rho_1^*}=\frac{6g}{(\rho_2-\rho_1^*)^4}-\frac{1}{{\rho_
1^*}^2}=0\;.
\end{equation}
The only solution between $\rho_2$ and the origin is 
\[\rho_1^*=\rho_2+\sqrt{\frac{3g}{2}}-\sqrt{\rho_2\sqrt{6g}+\frac{3g}{2}
}\;.\]
This is indeed a minimum since the second derivative of ${\cal N}$
with respect to $\rho_1$ is always positive when
$0\leq\rho_1\leq\rho_2$.

Substituting $\rho_1^*$ into ${\cal N}$ we find that
\begin{equation}
{\cal
N}\left(\rho_1^*\right)=\frac{1}{\rho_2+\sqrt{\frac{3g}{2}}-\sqrt{\rho_2
\sqrt{6g}+\frac{3g}{2}}}
-\frac{1}{\rho_2}
+\frac{2g}{\left(\sqrt{\rho_2\sqrt{6g}+\frac{3g}{2}}-\sqrt{\frac{3g}{2}}
\right)^3}
-\frac{2g\rho_3}{(\rho_2+\rho_3)(\rho_3-\rho_2)^3}\;.
\label{eq:Nstar}
\end{equation}
The first three terms of ${\cal N}\left(\rho_1^*\right)$ decay when 
$\rho_2$ is large. The fourth term however will remain finite unless
the difference $\rho_3-\rho_2$ diverges with $\rho_2$. If the fourth
term does remain finite, ${\cal N}\left(\rho_1^*\right)$ becomes
negative, the velocity of the first step becomes positive and the
first step cannot pass $\rho_1^*$ on its 
way to the origin. 

The above analysis indicates that the step density vanishes near
the facet edge. We consider two scenarios for the collapse of the
first step. In the first scenario the step starts the collapse from a
position below $\rho_1^*$. In this case the separation between the
first and second steps is initially divergent since $\rho_2-\rho_1^*$
diverges
as $\sqrt{\rho_2}$. But the initial configuration in the collapse of
one 
step is the final configuration of the former collapse period. Thus
the former period ended with a divergent distance $\rho_3-\rho_2$. 

In the second scenario the first step starts to 
collapse from a position above $\rho_1^*$. If the first step collapses
alone (i.e $\rho_2$ 
is large throughout the collapse period), the distance $\rho_3-\rho_2$
must grow to allow $\rho_1$ to pass through $\rho_1^*$. 

Either way there must be some point in the collapse period where the
separation $\rho_3-\rho_2$ diverges. At this point the step density
near the facet edge vanishes.

\end{document}